\documentclass[traditabstract]{aa} 

\usepackage{graphicx} 
\usepackage{setspace}
\usepackage[switch,pagewise]{lineno} 
\usepackage{rotating} 
\usepackage{natbib} 
\usepackage[usenames]{color} 
\usepackage{multirow} 
\usepackage{url} 
\usepackage{ulem} 
\usepackage{afterpage} 
\bibpunct{(}{)}{;}{a}{}{,} 
\newcommand{\kms} {{\rm ~ km \ s^{-1}}} 
\newcommand{\NHI} {N(\ion{H}{i})} 
\newcommand{\NHmol} {N(\mathrm{H_{2}})} 

\newcommand{\ratio} {\NHmol/\mathrm{I}_{\mathrm{CO(1-0)}}}
 
\newcommand{\Xunit} {~{\rm cm^{-2}/(K\kms)}} 
\newcommand{\mum} {{\rm Ê~ \mu m} }
\newcommand{\ICO} {${\rm I_{CO}}$ } 
\usepackage{txfonts}

\begin{document}
%\linenumbers
%
\title{Molecular and Atomic Gas in the Local Group Galaxy M33}
%   \title{Molecular Clouds in the outer disk of M33}
\titlerunning{Molecular and Atomic Gas in the Local Group Galaxy M33} \subtitle{}
\author{P.~Gratier \inst{1} \and J.~Braine \inst{1} \and N.J.~Rodriguez-Fernandez \inst{2} \and K.F.~Schuster \inst{2} \and C.~Kramer \inst{3} \and E.~M.~Xilouris \inst{4} \and F.~S.~Tabatabaei \inst{5} \and C.~Henkel \inst{5} \and E.~Corbelli \inst{6} \and F.~Israel \inst{7} \and P.~P. van~der~Werf \inst{7} \and D.~Calzetti \inst{8} \and S.~Garcia-Burillo \inst{9} \and A.~Sievers \inst{3} \and F.~Combes \inst{10} \and T.~Wiklind \inst{11}\and N.~Brouillet \inst{1} \and F.~Herpin \inst{1} \and S.~Bontemps \inst{1} \and S.~Aalto \inst{12} \and B.~Koribalski \inst{13} \and F.~van~der~Tak \inst{14} \and M.~C.~Wiedner \inst{10} \and M.~R\"ollig \inst{15} \and B.~Mookerjea \inst{16} }

\institute{Laboratoire d'Astrophysique de Bordeaux, Universit\'e de Bordeaux, OASU, CNRS/INSU, 33271 Floirac, France\\
\email{gratier@obs.u-bordeaux1.fr} \and IRAM, 300 Rue de la piscine, F-38406 St Martin d'H\`eres, France \and Instituto Radioastronomia Milimetrica (IRAM), 
   Av. Divina Pastora 7, Nucleo Central, E-18012 Granada, Spain
 \and Institute of Astronomy \& Astrophysics, National Observatory of Athens, I. Metaxa \& V. Pavlou, P. Penteli,
15236 Athens, Greece \and Max-Plank-Institut f\"ur Radioastronomy (MPIfR), Auf dem H\"ugel 69, 53121 Bonn, Germany \and INAF - Osservatorio Astrofisico di Arcetri, L.E. Fermi, 5 - 50125 Firenze, Italy \and Leiden Observatory, Leiden University, PO Box 9513,  NL 2300 RA Leiden, 
The Netherlands \and Department of Astronomy - LGRT, University of Massachusetts - Amherst, 710 North Pleasant Street, Amherst, MA 01002, USA \and Observatorio Astronomico Nacional (OAN)-Observatorio de Madrid, Alfonso XII, 
3, 28014-Madrid, Spain \and Observatoire de Paris, LERMA, CNRS, 61 Av de l'Observatoire, 75014 Paris \and Space Telescope Science Institute, 3700 San Martin Drive, Baltimore, MD 21218, USA \and Department of Radio and Space Science, Chalmers University of Technology, Onsala Observatory, SE-439 94 Onsala, Sweden \and CSIRO Astronomy \& Space Science, Australia Telescope National Facility, P.O. Box 76,  Epping NSW 1710, 
Australia  \and SRON Netherlands Institute for Space Research, Landleven 12, 9747 AD Groningen, The Netherlands \and 1.Physikalisches Institut, Universit\"at zu K\"oln, Z\"ulpicher Str. 77, 
D-50937 K\"oln, Germany \and Department  of Astronomy \& Astrophysics, Tata Institute of Fundamental Research, Homi Bhabha Road, Mumbai 400005, India
}

\date{}
\abstract{We present high resolution large scale observations of the molecular and atomic gas in the Local Group Galaxy M33. The observations were carried out using the HEterodyne Receiver Array (HERA) at the 30m IRAM telescope in the \mbox{CO(2--1)} line achieving a resolution of $12\arcsec \times 2.6\kms$, enabling individual Giant Molecular Clouds (GMCs) to be resolved. The observed region is 650 square arcminutes mainly along the major axis and out to a radius of 8.5 kpc, and covers entirely the $2\arcmin \times 40\arcmin$ radial strip observed with the HIFI and PACS 
Spectrometers as part of the {\tt HERM33ES} Herschel key program. The achieved sensitivity in main beam temperature is \mbox{20--50~mK} at $2.6~\kms$ velocity resolution. The \mbox{CO(2--1)} luminosity of the observed region is $1.7\pm0.1\times10^{7}~{\rm K\kms pc^2}$ and is estimated to be $2.8\pm0.3\times10^{7}~{\rm K\kms pc^2}$ for the entire galaxy, corresponding to H$_2$ masses of $1.9\times10^8~{\rm M_{\sun}}$ and  $3.3\times10^8~{\rm M_{\sun}}$ respectively (including He), calculated with a $\ratio$ twice the Galactic value due to the half-solar metallicity of M33. \ion{H}{i}~21~cm VLA archive observations were reduced and the mosaic was imaged and cleaned using the multi-scale task in the CASA software package, yielding a series of datacubes with resolutions ranging from $5\arcsec$ to $25\arcsec$. The \ion{H}{i} mass within a radius of 8.5 kpc is estimated to be $1.4 \times 10^9{\rm M_{\sun}}$. The azimuthally averaged CO surface brightness decreases exponentially with a scale length of $1.9\pm0.1$~kpc whereas the atomic gas surface density is constant at $\Sigma_{\ion{H}{i}}=6\pm2~{\rm M_{\sun}}$pc$^{-2}$
deprojected to face-on. 
For a $\ratio$ conversion factor twice that of the Milky~Way, the central kiloparsec H$_2$ surface density is ${\rm \Sigma_{H_{2}}=8.5\pm0.2~{\rm M_{\sun}}pc^{-2}}$. The star formation rate per unit molecular gas (SF~Efficiency, the rate of transformation of molecular gas into stars), as traced by the ratio of CO to H$_{\alpha}$ and FIR brightness, is constant with radius. The SFE, with a $\ratio$ factor twice galactic, appears 2--4 times greater than of large spiral galaxies. A morphological comparison of molecular and atomic gas with tracers of star formation is presented showing good agreement between these maps both in terms of peaks and holes. A few exceptions are noted. Several spectra, including those of a molecular cloud situated more than 8~kpc from the galaxy center, are presented.}


\maketitle

\section{Introduction}
\begin{table}
	\begin{minipage}
		{88mm} \caption{\label{tab.M33prop}Adopted parameters for M33} 
		\begin{tabular*}
			{88mm}{@{\extracolsep{\fill}}lr} \hline\hline\noalign{\smallskip} 
			$\alpha$(J2000) & $1^\mathrm{h}33^\mathrm{m}50\fs9$\\
			$\delta$(J2000) & $+30\degr39\arcmin39\arcsec$\\
			Distance & 840~kpc\,\,\footnotemark[1]\\
			Optical Radius {\rm $R_{25}$} & $30.8\arcmin$\,\, \footnotemark[2]\\
			Inclination & $56\degr$\,\, \footnotemark[2]\\
			Position angle & $22.5\degr$\,\, \footnotemark[2]\\
			Central Oxygen abundance & $12 + \log(O/H) = 8.4$\,\, \footnotemark[3]\\
			%1500 \AA& $5\arcsec$ & GALEX FUV \footnotemark[4]\\
			\noalign{\smallskip}\hline 
		\end{tabular*}
		\footnotetext[1]{\citet{Galleti.2004}} \footnotetext[2]{{\tt HYPERLEDA} \citep{Paturel.2003}} \footnotetext[3]{\citet{Magrini.2009}}
	\end{minipage}
\end{table}

The Local Group galaxies span a broad range in mass, luminosity, morphology, and metallicity.  Two large spirals (the Milky 
  Way and M31) are the centers of two galaxy sub-groupings, each being
  surrounded by a large number of dwarf galaxies. In addition, M31 --- the 
  Andromeda Galaxy --- has a small spiral companion, M33 (the Triangulum Galaxy);
  their separation is approximately 15 degrees, corresponding to 200 kpc (assuming a common
  distance of 840 kpc; \citet{Galleti.2004}). Gaseous streams are observed 
  between them, indicating tidal interaction \citep{Putman.2009}. 
   %Of these, with the Milky~Way,  M31 is the only other large spiral and M33 is the only small spiral, the others being irregular dwarf or spheroidal galaxies. 
   M33 provides a means of observing a galaxy morphologically similar to our own but with a mass only a tenth of the Milky~Way and factor two lower metallicity {  \citep{Rosolowsky.2008,Magrini.2009}}. Further evidence for the difference between M33 and the Milky~Way
is the large gas fraction and blue stellar colors of the former relative
to the latter. M33 thus represents an environment in which to study the interstellar medium (ISM) and star formation (SF) that cannot be replaced by Galactic observations and where individual GMCs can be resolved to probe their star formation activity. It may also be possible to apply what we learn by studying M33 to the physics of early-universe objects, which share many of the characteristics of M33.

In this article we present sensitive and high-resolution mapping observations of the CO $J = 2 \rightarrow 1$ transition in M33 in order to study the morphology and dynamics of the molecular component. The total mapped area covers 650 square arcminutes, mainly along the major axis of the galaxy. 

\begin{figure}
	[tbp]
	\begin{flushleft}
		\includegraphics[angle=0,width=8.8cm]{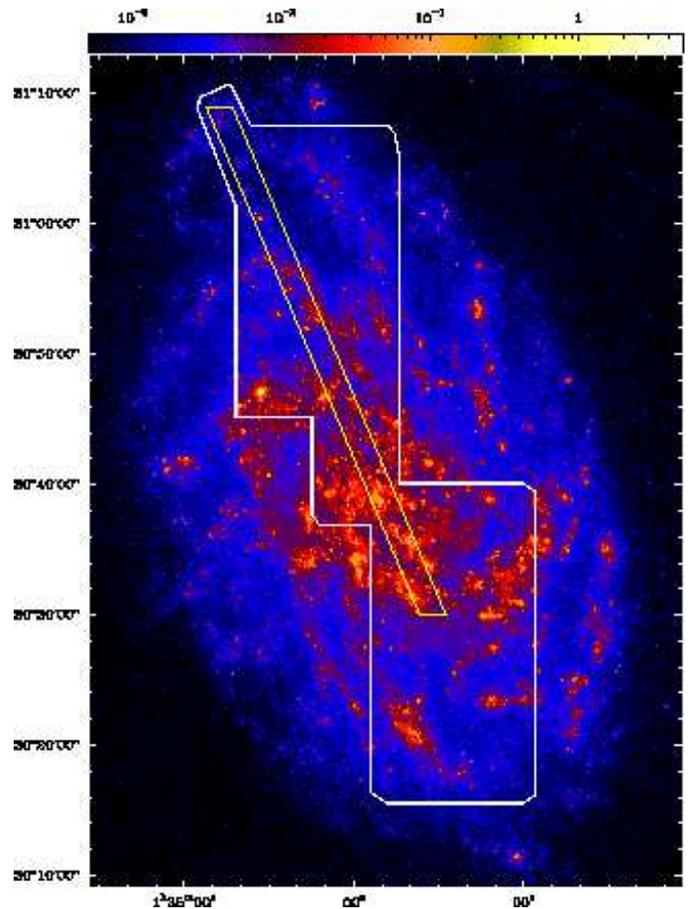} \caption{\label{fig.FUV_COcover}The Local Group galaxy M33. This color image shows the GALEX FUV data which trace young stars and dust in the disk through attenuation. Overlaid are the observed fields (a)  using the IRAM 30m Pico Veleta telescope in the CO(2--1) line (thick white outline) and (b) by HIFI/PACS instruments as part of the {\tt HERM33ES} Herschel Key Program  (thin yellow stripe).
  %In white: area covered by our CO(2--1) map overlaid on \emph{GALEX} FUV data. In yellow: 2$\arcmin$ wide strip along the major axis that will be observed by HIFI instrument as part of the {\tt HERM33ES} Herschel Key Program 
  }  
	\end{flushleft}
\end{figure}

%In a companion paper, a large sample of Giant Molecular Clouds (GMC) detected in these observations will be described (É)
A $2\arcmin \times 40\arcmin$ wide strip along the major axis (see Fig.~\ref{fig.FUV_COcover}) was observed to a particularly low noise level  of 25~mK at $2.6~\kms$ velocity resolution to compare with the sensitive \ion{C}{ii} Herschel/HIFI and Herschel/PACS spectroscopy observations which will be obtained as part of the {\tt HERM33ES} Herschel Key Program \citep{Kramer.2010}. While the most sensitive and among the highest resolution, these are not the first maps of M33 in the CO lines. \citet{Engargiola.2003} observed the whole of the inner disk (up to about 5kpc along the major axis) with the BIMA array at $13\arcsec$ resolution; \citet{Heyer.2004} observed the inner disk and a small major axis strip at 50$"$ resolution with FCRAO; \citet{Rosolowsky.2007a} combined the BIMA$+$FCRAO$+$NRO data to improve the sensitivity and resolution of the previous maps; and \citet{Gardan.2007} observed a rectangle at high sensitivity and $15\arcsec$ resolution extending from NGC~604 to the R$_{25}$ radius. {  Table \ref{tab.surveys} summarizes the characteristics of previous molecular and atomic gas surveys in M33}. This work extends the \citet{Gardan.2007} work further North and to the South at higher resolution and sensitivity. Earlier studies of GMCs in M33 include \citet{Wilson.1997} and \citet{Rosolowsky.2003} and studies similar to our own of other Local Group galaxies have been made by e.g. \citet[][]{Fukui.2008,Israel.2003,Leroy.2006,Nieten.2006}.

As mentioned by \citet{Blitz.2006} for M33 and \citet{Leroy.2006} for IC~10 and discussed more extensively by \citet{Gardan.2007}, the Star Formation Rate (SFR) per unit H$_2$ mass or Star Formation Efficiency (SFE $=$ SFR${\rm /M(H_2)}$ in~yr$^{-1}$) was found to be up to an order of magnitude higher in these small galaxies than in large spirals. This appears to be the case in distant galaxies as well, given the factor 10 increase in (commoving) SFR density \citep[e.g.][]{Madau.1996, Wilkins.2008}. Are there local universe analogs of these distant objects? Is M33 one of them?

One of the obvious questions is whether the H$_2$ mass has not been underestimated in these subsolar metallicity objects. The articles using the data presented here and as part of the {\tt HERM33ES} project will attempt to answer that issue clearly. %Assuming the H$_2$ mass estimate is appropriate, what is the cause of the quick conversion of H$_2$ into stars? A further possibility might be that the stellar initial mass function (IMF) in M33 is top heavy, causing the SFR (and SFE) to be overestimated -- but then what changes the IMF?

%The metallicity gradient in M33 is very weak, unlike most spirals, and we assume $12 + log(O/H) = 8.4-0.03 R_{~kpc}$ \citep{Rosolowsky.2008, Magrini.2009} when necessary. 

Metallicities were lower in the past and H$_2$ production is believed to take place on grain surfaces --  therefore \ion{H}{i} to ${\rm H_2}$ conversion is expected to be less efficient in
low metallicity systems \citep{Krumholz.2009a}, and this is indeed observed in many systems \citep[e.g.][]{Leroy.2007}.
%therefore \ion{H}{i} to H$_2$ conversion is expected to be less efficient in younger systems and indeed in the local universe the low-Z objects have lower H$_2$/\ion{H}{i} fractions. 
The conversion of \ion{H}{i} to H$_2$ in the intermediate redshift systems would have to be much more efficient than today, generating unusually high molecular fractions in these distant galaxies (contrary to expectations) with less stellar gravity, in order to have a similar efficiency in converting H$_2$ to stars. 

Here, we will present new maps of CO and HI. In this first paper, we restrict ourselves to a study of the radial distribution of the molecular gas and infrared surface brightness, the molecular and atomic gas surface densities and the star formation efficiency and to a qualitative comparison between the maps of star formation rate tracers, i.e. the dust maps, and the gas maps. In addition,
we will discuss CO spectra in a few selected regions. A series of articles will follow, focussing on at least ($i$) cloud populations, life cycle, and mass spectrum  ($ii$) dynamics of the molecular gas and the role of spiral arms  ($iii$) diffuse CO emission, after subtraction of the clouds identified and  ($iv$) a more detailed comparison of the star formation rate -- gas surface density relation. 
%Rather, we think that it is more likely that the initial conditions when the H$_2$ is formed are such that the subsequent collapse can happen more quickly and/or create stars with a higher-mass IMF (the tracers of star formation are biased towards high-mass stars).

\section{Observations and data reduction} 
\begin{table*}
	[tbp] \caption{\label{tab.surveys}Comparison of past survey characteristics.}
	\begin{minipage}{\textwidth}  
	\begin{flushleft}
		\begin{tabular*}
			{\textwidth}{@{\extracolsep{\fill}}lrrrrrrr} \hline\hline\noalign{\smallskip} 
			CO Survey & Telescope& Line & Angular    & Spectral   & Area & R$_{max}$ & Sensitivity \\
			          &          &      & resolution & resolution &      &           & per channel             \\
			          &          &      & ($\arcsec$) & ($\kms$) &   (arcmin$^2$)   &  (kpc)         &        (mK)      \\
			\noalign{\smallskip}\hline\noalign{\smallskip} 
			This paper & IRAM 30m & CO(2--1) & $12$ & $2.6$ &$643$ &8 & 15--30\\
			\citet{Gardan.2007}&IRAM 30m & CO(2--1) & $15$&$2.6$ & $253$ & 7& 15--30\\
			\citet{Rosolowsky.2007a}& BIMA+NRO+FCRAO & CO(1--0) & $20$ & $2.6$ & 170 & 2 & 60\\
			\citet{Rosolowsky.2007a}& BIMA+FCRAO & CO(1--0) & $13$ & $2$ & $900$ & 5.5 & 240\\
			\citet{Heyer.2004} &FCRAO & CO(1--0) & 45 & 1 & 900 & 5.5 & 53\\
			\citet{Engargiola.2003} &BIMA & CO(1--0) & $13$ &$2$& 900 &   5.5& 240\\
			\end{tabular*}
			\begin{tabular*}{\textwidth}{@{\extracolsep{\fill}}lrrrrrr} \noalign{\smallskip}\hline\hline\noalign{\smallskip} 
			HI Survey & Telescope&  Angular    & Spectral    & R$_{max}$ & Sensitivity& Sensitivity        \\
			          &          &  resolution & resolution   &           & per channel& per channel        \\
			          &          & ($\arcsec$) & ($\kms$)   &  (kpc)        &        (mJy/beam)   &        (K)      \\
			
			\noalign{\smallskip}\hline\noalign{\smallskip}
			This paper & VLA BCD & 5--25\footnotemark[1] & 1.3 & 9 & 1.1--2.8\footnotemark[1] & 24--2.75\footnotemark[1]\\
			\citet{Putman.2009} & Arecibo & 204 & 5.15 &  22   & 55 &0.3\\
			\citet{Corbelli.1997} & Arecibo&  270\footnotemark[2]  & 4.1&  20&$\ldots$\footnotemark[3] &$\ldots$\footnotemark[3]\\
			\citet{Deul.1987} & WRST+Effelsberg 100m & $12 \times 24$ & 8.2 & 9 & $\ldots$ & 1.2\\
			\noalign{\smallskip}\hline
		\end{tabular*}
		\footnotetext[1]{See Table~\ref{tab.HIcubes} for details.}
		\footnotetext[2]{Undersampled hexagonal grid, the value corresponds to the grid spacing.}
		\footnotetext[3]{The authors give an integrated intensity sensitivity value of 1--2~Jy$\kms$ or 5.5--11~K$\kms$.}
	\end{flushleft}
	\end{minipage}
\end{table*}

\subsection{\label{sec.CO} IRAM CO(2--1) observations} The observations presented here are a follow-up to the \citet{Gardan.2007} mapping of a large part of M33. All data were taken with the 30 meter telescope run by the Institut de RadioAstronomie Millim\'etrique\footnote{IRAM is supported by INSU/CNRS (France), MPG (Germany) and IGN (Spain).} (IRAM) on Pico Veleta near Granada, Spain. The observations were carried out starting in August 2008 and also include the data taken by \citet{Gardan.2007}.

The IRAM HERA instrument (HEterodyne arRAy), composed of 9 dual-polarization 1.3~mm receivers \citep{Schuster.2004}, was used in On-The-Fly scanning mode. All data are presented in the main beam temperature scale and we have assumed forward and main beam efficiencies of $\eta_{for} = 0.90$ and $\eta_{mb} = 0.49\pm0.04$ for the HERA observations, the sensitivity is then ${\rm 9.6 Jy/K}$ \citep{Schuster.2004}. The scanning speed was 3 arcseconds per second of time with dumps every second. Both the VESPA backend, at 1.25MHz resolution, and the WILMA correlator with 2MHz channel spacing, were used simultaneously. Reference positions were observed roughly every three minutes and these were taken to be holes in the \ion{H}{i} column density outside of the M33 disk with no visible FIR emission. The average beam size over the nine pixels is $11.7\arcsec$. The array was tilted by 18.5 degrees, obtaining 7.6 arcsecond spacing between the individual pixel tracks, and then shifted in position by 3.8 arcsec in order to obtain full sampling. This was first done for scanning parallel to RA, then parallel to Dec, the array was then rotated by 90 degrees and the procedure was repeated. The array rotation is to have different pixels cover the same regions to improve data homogeneity and reduce striping. For the HIFI low noise major axis strip, additional scans along and perpendicular to the strip direction where acquired. The repetition of independent scans over the same areas also randomizes any pointing errors. Thus, if there are pointing uncertainties of 2$\arcsec$, these contribute negligibly to the overall beam size (through a negligible smearing). The part of M33 presented here comes from the observation of about 5 million spectra.

Reduction was carried out within Gildas\footnote{\url{http://www.iram.fr/IRAMFR/GILDAS}} CLASS and GREG software. The intensity of the CO lines is usually small so that emission is not seen in individual spectra, the baseline removal was therefore implemented using a windowing based on the \ion{H}{i} data at $17\arcsec$ presented in section~\ref{sec.HI}. In a first step a linear baseline is fitted and subtracted from each individual spectrum not taking into account channels inside a window computed by finding the first channels at a $3\sigma$ level framing the peak channel of \ion{H}{i} emission and adding a $40\kms$ buffer on each side of this. 
A filtering step is then applied to remove spectra with anomalous noise. The actual noise is computed for each spectrum and compared to the theoretical noise computed from the system temperature, integration time and spectral resolution. The spectra with excess noise attributable to poor baselines are removed; this corresponds to about 10\% of the spectra taken.
The remaining spectra are then convolved by a gaussian to obtain a regularly gridded cube of the desired resolution. We then fit and remove a third order polynomial to the spectra corresponding to each of the spatial pixels in the gridded cube this time reducing the velocity buffer described previously to $30\kms$. With this method we have made CO cubes with spacial resolutions of 12, 15 and 25$\arcsec$ and with velocity resolutions of $2.6\kms$.

\begin{figure}
	[tbp] 
	\begin{flushleft}
		\includegraphics[angle=0,width=8.8cm]{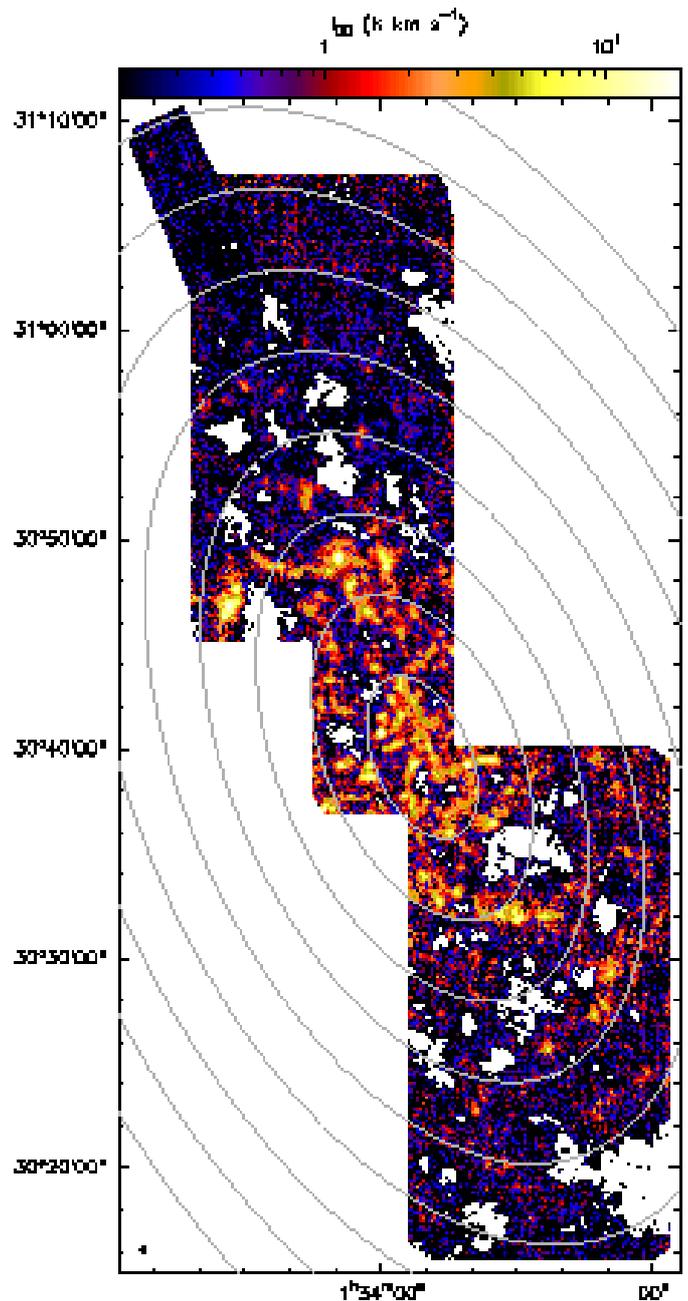} \caption{\label{fig.COmom0}IRAM CO(2--1) integrated intensity map of the galaxy M33 at a resolution of $12\arcsec$.
                                 This map was made using the VLA \ion{H}{i} cube ($17\arcsec$ resolution) as a mask, employing a
                                 10~K cutoff per channel (see Sect~\ref{sec.HI}). As a consequence, areas without HI emission above that
                                 level appear white. Grey ellipses are drawn every kpc using orientation parameters
                                 listed in Table~\ref{tab.M33prop}.} 
	\end{flushleft}
\end{figure}

\begin{figure}
	[tbp] 
	\begin{flushleft}
		\includegraphics[angle=0,width=8.8cm]{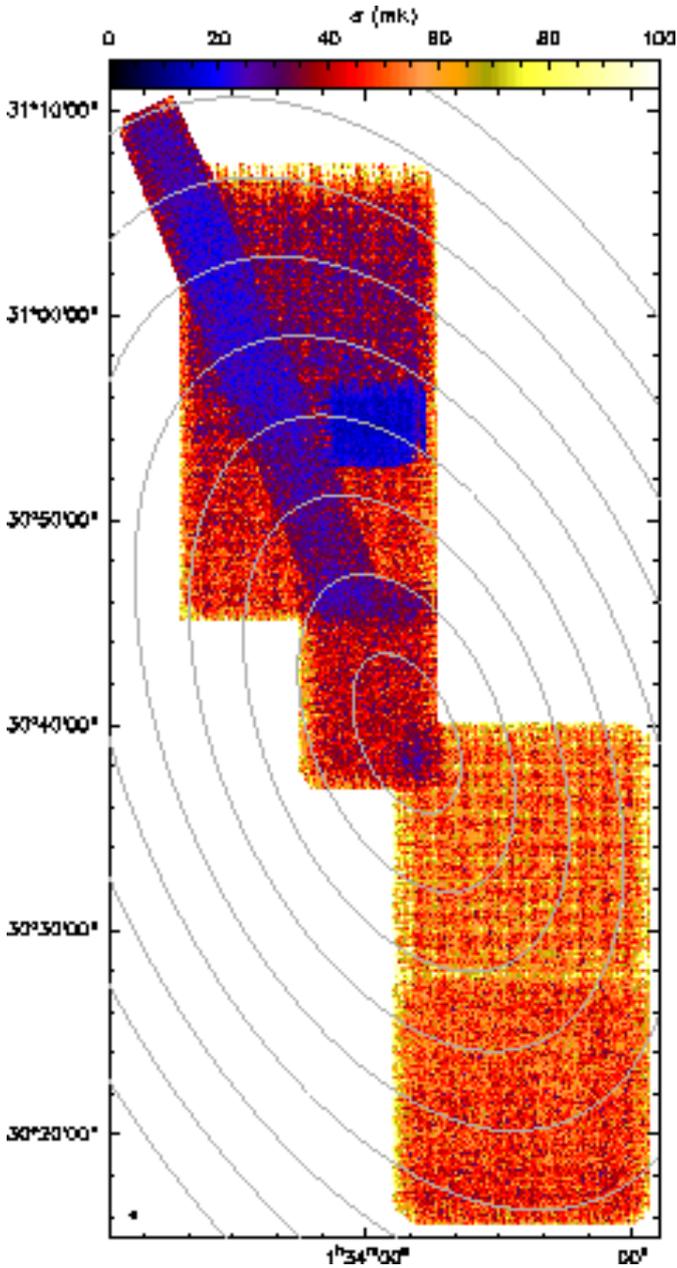} \caption{\label{fig.COnoise}CO noise map of the 12$\arcsec$ cube computed from 30 channels free of M33 emission, from $-80$ to $0\kms$ {$\rm V_{LSR}$}.}  
	\end{flushleft}
\end{figure}

We compute the CO(2--1) integrated intensity map using a masking method \citep{Gratier.2010} taking into account \ion{H}{i} data, we developed in order to filter out some of the noise present in the observations and increase the sensitivity to low intensity possibly diffuse CO emission. Previous masking methods used masks created from spatially smoothed versions of the original CO data cubes to filter out regions dominated by noise \citep{Adler.1992,Digel.1996,Loinard.1999}. We use the 21~cm atomic hydrogen data at $17\arcsec\times17\arcsec \times 1.27\kms$ resolution presented in section~\ref{sec.HI} to achieve the same goal, the underlying hypothesis being that molecular gas is unlikely to be present for low enough values of $\NHI$ so the corresponding velocity channels can be discarded when computing the integrated intensity CO map. For each pixel of the \ion{H}{i} cube, we create a binary mask keeping only the velocity range for each pixel corresponding to a \ion{H}{i} signal value above a defined threshold in the \ion{H}{i} signal. The integrated moment map for the CO(2--1) data is then computed summing only velocity channels included in the \ion{H}{i} mask. Figure~\ref{fig.COmom0} shows the integrated intensity map for the $12\arcsec$ cube using this method and Fig.~\ref{fig.COnoise} corresponding noise map at 12$\arcsec$ computed from 30 channels free of emission from M33 (from $-80$ to $0\kms$ {\rm $V_{LSR}$}).
The result is an increased S/N ratio as the channels contributing only noise to the sum are no longer taken into account. The value of the \ion{H}{i} signal threshold was chosen at 10~K which corresponds to $1.9\sigma$ where $\sigma$ is the {\it rms} noise level in the $17\arcsec$ \ion{H}{i} cube. We tested masking values between 5 and 25~K (0.9 to $4.6\sigma$) and the total CO intensity varied by only a few percent. Significantly above or below these values, CO signal is lost or more noise is included. At the 10~K brightness level, the \ion{H}{i} lines are broad and sample several channels to either side of the rotation curve. Comparing with other techniques of looking for CO, we found no evidence of missing CO emission by this technique. As an example, the ``Lonely Cloud'' ($\alpha=1^\mathrm{h}34^\mathrm{m}16\fs7$, $\delta=+30\degr59\arcmin3\arcsec$, J2000) \citep{Gardan.2007} in an interarm region not near an \ion{H}{i} maximum is included by this method.  
\subsection{\label{sec.HI}VLA \ion{H}{i} mosaicing} 
\begin{figure}
	[tbp] 
	\begin{flushleft}
		\includegraphics[angle=0,width=8.8cm]{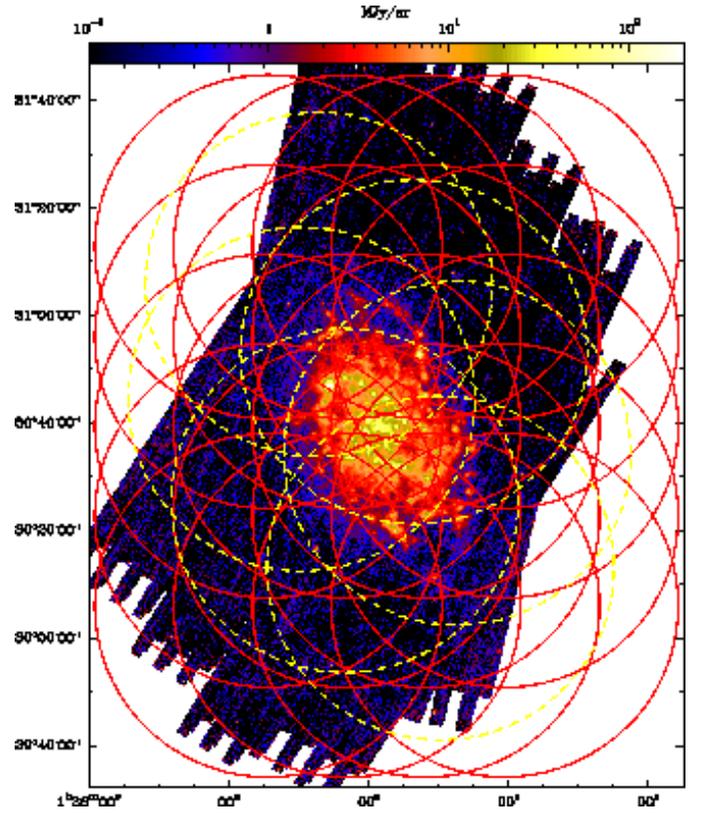} \caption{\label{fig.HIcover} \ion{H}{i} primary beam coverage of M33 overlaid on the $70\mu$m \emph{Spitzer} map. VLA D pointings are in red (solid lines) and B/C ones in yellow (dashed lines).} 
	\end{flushleft}
\end{figure}

The \ion{H}{i} observations are from archival VLA B, C, and D array data taken as part of projects AT206 and AT268 in 1997, 1998 and 2001.  At the frequency of the \ion{H}{i} line the VLA primary beam is 32\arcmin. The mosaic comprises 20 D array pointings on a nearly square grid separated by 16\arcmin   and 6 B and 6 C pointings in a diamond shape covering the entire star forming disk. The primary beam centers of the B/C and D array mosaic are shown in Fig.~\ref{fig.HIcover} overlaid on the \emph{Spitzer} $70\mu$m map -- the D array data contains more pointings extending further from the center of M33. Since our goal is compare the distribution and kinematics of the atomic and molecular gas at high resolution, we focus here on the stellar disk where VLA B-, C- and D-array data are available. 
Data calibration and imaging were carried out using the 2.4 version of the CASA\footnote{\url{http://casa.nrao.edu/}} software package. The quasar 3C48 ($\alpha=1^\mathrm{h}37^\mathrm{m}41\fs3$, $\delta=+33\degr09\arcmin35\arcsec$) was used as flux, bandpass and phase calibrator (with an adopted flux of 16.04~Jy). The calibration of the visibilities was done through the following tasks. First the correct flux was applied to the calibrator visibility model using {\tt setjy}, then frequency bandpass and gain calibrations were computed using the {\tt bandpass} and {\tt gaincal} tasks. 
 
\begin{figure}
	[tbp] 
	\begin{flushleft}
		\includegraphics[angle=0,width=8.8cm]{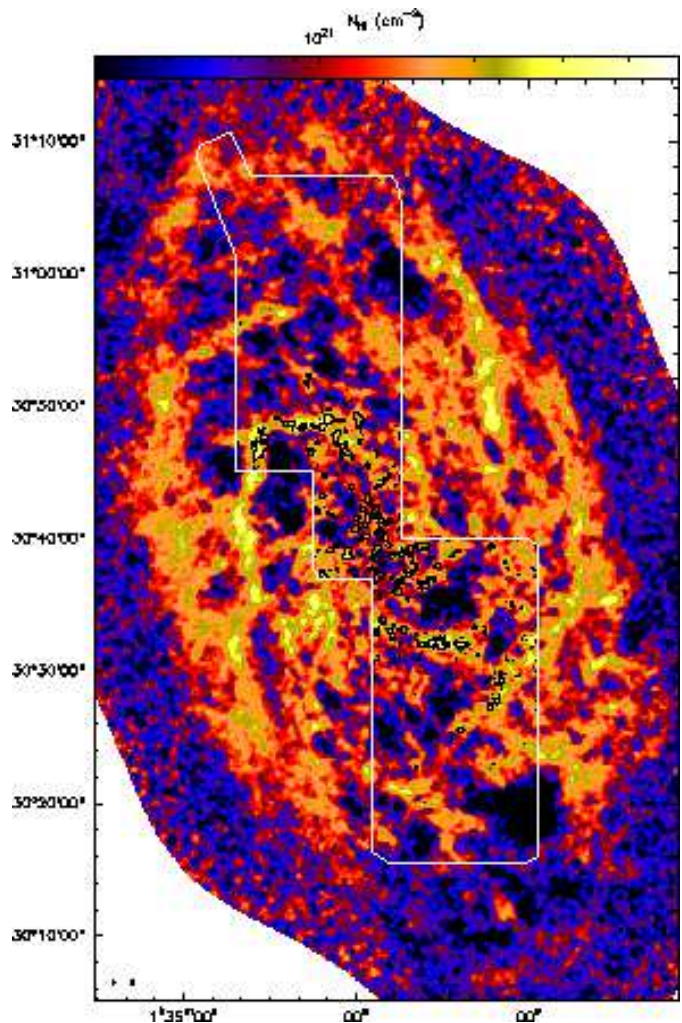} \caption{\label{fig.HImom0} VLA HI column density map (in color) of the galaxy M33 at a resolution of $17\arcsec$ overlaid
                                  with the IRAM \mbox{CO(2--1)} intensity contour at 1~K$\kms$. The region observed in \mbox{CO(2--1)}
                                  are marked with thin white lines. }  
	\end{flushleft}
\end{figure}

The calibration solutions were finally applied to the entire dataset with the use of the {\tt applycal} task, continuum emission averaged on channels not including line emission from M33 was subtracted in the \emph{uv}-space using the {\tt uvcontsub} task. Flagging of bad data was done by hand. The imaging was achieved using the multi-scale clean {\tt MSCLEAN} algorithm described in \citet{Cornwell.2008}. This method has several advantages over the classical clean algorithms that use point like sources to model the observed emission. By cleaning several scales simultaneously, it is able to efficiently recover extended emission and eliminates problems of flux correction \citep{Jorsater.1995}, negative bowls surrounding regions of extended emission, and the pedestal of low level uncleaned flux in the final residual map. An extensive comparison of multiscale clean and its point source counterpart has been carried out in \citet{Rich.2008}. A series of four cubes with angular resolutions of 5, 12, 17, and 25 arcseconds and  $1.27\kms$ channel width were computed. The 5$\arcsec$ cube corresponds to the highest resolution achievable with the B array. The other cubes were obtained using gaussian tapers in the \emph{uv}-plane of respectively 4900 and 3150 and 1900 meters FWHM. In order to minimize sidelobes over the entire map, a robust weighting scheme \citet{Briggs.1995}  was applied to the data using a robustness factor of $0.5$. The rms noise for the cubes was computed over the channels without emission from M33 within a galactocentric radius of 8.5 ~kpc. Table~\ref{tab.HIcubes} summarizes the beam sizes and noise properties of the different cubes. A zeroth moment map (Fig.~\ref{fig.HImom0} with the CO 1~K~$\kms$ intensity contour superposed) was computed at 17$\arcsec$ resolution by masking out regions of the $17\arcsec$ cube where emission from the 25$\arcsec$ cube was below 2.75K ($1\sigma$). The data presented here will be made available throught the Centre de Donn\'{e}es de Strasbourg (CDS)\footnote{\url{http://cdsweb.u-strasbg.fr/}}.
\begin{table}
	[tbp] \caption{\label{tab.HIcubes}Beam and rms noise properties of the \ion{H}{i} 21~cm datacubes.}  
	\begin{flushleft}
		\begin{tabular*}
			{88mm}{@{\extracolsep{\fill}}lrrr} \hline\hline\noalign{\smallskip} 
			Beam & PA & $\sigma_{S}$ & $\sigma_{T}$\\
			($\arcsec \times \arcsec$) & ($\degr$) & (mJy/beam) & (K)\\
			\noalign{\smallskip}\hline\noalign{\smallskip}
			 $5.5 \times 5.2$ & -95.1& 1.1& 24\\
			$12.0 \times 11.6$ & -31.8& 2.0& 9\\
			$17.2 \times 17.1$ & -45.8& 2.5& 5.4\\
			$25.9 \times 24.2$ & -74.8& 2.8& 2.75\\
			\noalign{\smallskip}\hline 
		\end{tabular*}
		The {\it rms} noise, in mJy/beam and brightness temperature, was calculated over an ellipse of galactocentric radius 8.5~kpc after primary beam correction. 
	\end{flushleft}
\end{table}

\section{Results}
\begin{table*}
	\begin{minipage}
		{\textwidth} \caption{\label{tab.avec_comp}Comparison of results with previous studies of M33.
		} 
		\begin{tabular*}
			{\textwidth}{@{\extracolsep{\fill}}lrrrr}
			\hline\hline\noalign{\smallskip}
			CO surveys     &   Molecular gas mass\footnotemark[2]    & SFR\footnotemark[7]        &  Depletion time\footnotemark[11] & Depletion time\footnotemark[12] \\
							&    (M$_{\sun}$)        & (M$_{\sun}$ yr$^{-1}$)                      &  ($10^8$ yr)\  & ($10^8$ yr)\\
			\noalign{\smallskip}\hline\noalign{\smallskip}
			This study                         & $1.9\times10^8$ ( $3.3\times10^8$)\footnotemark[3]  &0.27\footnotemark[8]   & 7   & 7        \\
			\citet{Gardan.2007}                & $3.2\times10^7$\footnotemark[4]                     &0.05--0.13\footnotemark[9] & 2--6 & 1--3\\
			\citet{Rosolowsky.2007a}           & $3.6\times10^8$\footnotemark[4]                     &$0.45\footnotemark[10]$  &  8 &  $\ldots$\\ 
			\citet{Heyer.2004}                 & $3.5\times10^8$\footnotemark[5]                      &$0.45\footnotemark[10]$ &  8  &1--3\\
			\citet{Engargiola.2003}            & $9\times10^7$\footnotemark[4]                       &$0.24\footnotemark[9]$  &   4  &$2$\\
			\citet{Corbelli.2003}\footnotemark[1]& $2.9\times10^8$\footnotemark[6]                     &$0.45\footnotemark[10]$  & 6  &$\ldots$\\
			\noalign{\smallskip}\hline\hline\noalign{\smallskip}
           HI surveys &Atomic gas mass\footnotemark[13]\\
           \noalign{\smallskip}\hline\noalign{\smallskip}
           This study & $1.4\times10^9$\\
           \citet{Putman.2009} &$1.5\times10^9$\footnotemark[14]\\
           \citet{Corbelli.1997}&$2.6\times10^9$\footnotemark[15]\\
           \citet{Deul.1987}&$1.1\times10^9$\\
           \noalign{\smallskip}\hline\noalign{\smallskip}
		\end{tabular*}
		\footnotetext[1]{FCRAO CO(1--0) data from \citet{Heyer.2004}.}
		\footnotetext[2]{All values converted to our adopted X${\rm _{CO}}=4\times10^{20}~\Xunit$ (twice Galactic).}
		\footnotetext[3]{The value in parentheses is the extrapolation to the whole galaxy for comparison with other surveys. See Sect. \ref{sec.H2mass} for details.}
		\footnotetext[4]{Original value used X${\rm _{CO}}=2\times10^{20}~\Xunit$ (Galactic).}
		\footnotetext[5]{Original value used X${\rm _{CO}}=3\times10^{20}~\Xunit$.}
		\footnotetext[6]{Original value used X${\rm _{CO}}=2.8\times10^{20}~\Xunit$.}
		\footnotetext[7]{Over the areas mapped in CO for each survey.}
		\footnotetext[8]{From \citet{Verley.2009}.}
		\footnotetext[9]{From H$_{\alpha}$ with a \citet{Kennicutt.1998} calibration.}
		\footnotetext[10]{We give the value from \citet{Engargiola.2003} as the CO mapped areas are identical.}
		\footnotetext[11]{Value computed by dividing the first by the second column of this table.}
		\footnotetext[12]{Original value given by the cited papers.}
		\footnotetext[13]{All values converted to our adopted distance ${\rm D}=840$~kpc.}	    
		\footnotetext[14]{Original value used a distance of 730~kpc. Inside a $3.2\times10^{20}$~cm$^{-2}$ contour.}
		\footnotetext[15]{Original value used a distance of 690~kpc. Over their whole mapped area extending beyond 8.5~kpc.}
	\end{minipage}
\end{table*}
\subsection{Gas masses}
In this section we present measurements of total integrated flux and derived total molecular and atomic gas masses.
\subsubsection{\label{sec.H2mass}Molecular Gas}
 The CO(2--1) luminosity measured over our observed area is $1.7\pm0.1\times10^{7}~{\rm K\kms pc^2}$. Our CO mapped region covers 28\% of the total galaxy area up to 8.5~kpc but $61\pm2\%$ of the infrared emission (see details in Sect. \ref{sec.radial}). The estimated CO(2--1) luminosity for the whole ${\rm R_{25}}$ disk is thus $2.8\pm0.3\times10^{7}~{\rm K\kms pc^2}$. The molecular gas mass is computed with the following hypotheses. Given that metallicity is half solar and the gradient in M33 is weak ($12 + log(O/H) = 8.4-0.03 R_{kpc}$, \citealp{Rosolowsky.2008, Magrini.2009}), we use a constant ``standard'' Milky~Way factor \citep{Dickman.1986} multiplied by a factor of two $\ratio = 4 \times 10^{20}\Xunit$,  implicitly assuming an inverse linear dependence between the conversion factor and metallicity \citep{Wilson.1995} but see \citet{Israel.2000}. 
%Other sources of variation of the conversion factor such as the gas pressure {  or non linear dependence with metallicity and interstellar radiation field intensity \citep{Israel.2000}} are not included in {  this} study.
Using the CO(1--0) observations from \citet{Rosolowsky.2007a} and our CO(2--1) observations we compute a {{CO(2--1)/C0(1--0)}} ratio of 0.8 in the central kiloparsec of M33.
%\citet{Sawada.2001} give a \mbox{CO(2--1)}/(1--0) line ratio of 0.96 for the inner kiloparsec of the Milky~Way.
Due to the decrease of the excitation temperature with radius we can expect lower values of the line ratio for the outer parts as found by \citet{Sawada.2001} for the Milky Way and \citet{Braine.1997} for NGC4414. With a linear gradient of the line ratio from 0.8 at the center of the galaxy to 0.5 at 8.5~kpc, and taking into account the measured exponential decrease of the CO emission with radius (see Sect~\ref{sec.radial})  we find an average \mbox{CO(2--1)}/(1--0) line ratio of 0.73 for the whole galaxy. The computed mass includes helium by multiplying by a factor 1.37. The molecular gas mass for the whole galaxy is thus $3.3\times10^8~{\rm M_{\sun}}$.
\subsubsection{Atomic Gas}
The integrated \ion{H}{i} flux density over a region corresponding to a galactocentric radius smaller than 8.5~kpc, slightly over the R$_{25}$ radius for M33, is ${\rm 8.4\times10^3~Jy~\kms}$. The atomic gas mass was derived from the \ion{H}{i} integrated intensity using a conversion factor $1.8\times10^{18}\Xunit$ \citep{Rohlfs.1996}. The total \ion{H}{i} mass over the same region is $1.4\times10^9~$M$_{\sun}$. 
%\citet{Corbelli.1997} using single-dish data from the Arecibo 305m dish find a total \ion{H}{i} mass (correcting their chosen distance to M33 (690~kpc) to match ours (840~kpc)) for the region of radii smaller than R$_{25}$ of $2 \times10^{9}$ M$_{\sun}$. 
 \citet{Putman.2009} using the Arecibo telescope find give $1.5\times10^9~$M$_{\sun}$ inside a very similar contour which they define as the star forming disk. %Using the same VLA data with short spacings from the Green Bank Telescope \citep[][private communication]{Thilker:private} find $1.44\times10^9$M$_{\sun}$ for the same region.
\subsection{Kinematics}
\begin{figure*}
	[htbp] 
	\begin{flushleft}
		\includegraphics[angle=270,width=180mm]{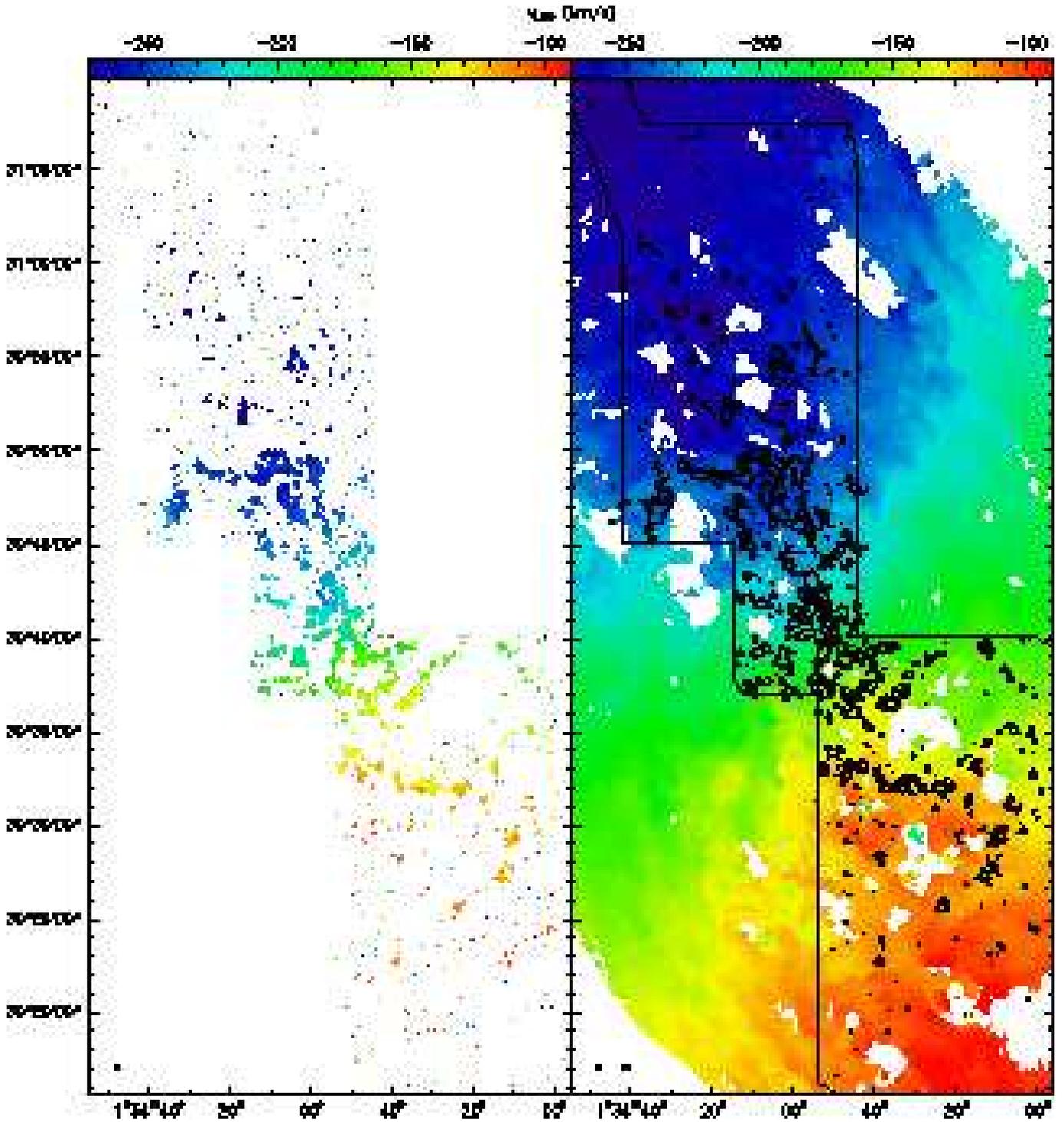} \caption{\label{fig.mom1}CO (\emph{left}) and \ion{H}{i} (\emph{right}) velocity fields at respectively $12\arcsec$ and {  17$\arcsec$}. The CO velocity field was computed from the original non \ion{H}{i} masked cube, both velocity fields are thus independent. The contours on the right correspond to a constant S/N ratio of 10 in the $12\arcsec$ CO integrated intensity.}  
	\end{flushleft}
\end{figure*}
{  We obtain velocity fields for both datasets by computing the first moment of the emission. In the case of \ion{H}{i}, the $17\arcsec$ data was masked by the $25\arcsec$ cube with a threshold of 10K.} For the CO data, we computed the first moment of the original non \ion{H}{i} masked datacube, masking out emission below $4\sigma$. Both methods are thus completely independent of each other. Fig.~\ref{fig.mom1} shows the CO and \ion{H}{i} velocity fields at respectively $12\arcsec$ and {  17$\arcsec$}.  
The excellent agreement between the two velocity fields is a further argument in favor of using the \ion{H}{i} masking to obtain CO integrated intensity maps.
%It will be seen in Section 6 that both a Tpeak and moment-based measurement of the velocity have their drawbacks due to double-peaked spectra that are present even at high resolution.
\subsection{\label{sec.radial} Radial distribution} 

The radial distribution of properties were derived taking azimuthal averages over elliptical rings with inclinations of i=$56\degr$ and position angle PA=$22.5\degr$ (measured toward East from North). The quantities are therefore deprojected from the sky plane onto the plane of M33. The radial step is 500pc which corresponds to about 2$\arcmin$. The bottom part of Fig.~\ref{fig.S_radius} shows the cumulative fractions of the area (solid line) and 70$\mum$ surface brightness (dashed line) in our CO map {  (black area in the inset)} compared to the whole elliptical region of radius less than R { (sum of grey and black areas of the inset).} For galactocentric distances below 2~kpc our map samples more than 80\% of the accessible disk at these radii both in area and 70$\mum$ flux. The whole CO mapped area corresponds to 28\% of the area up to 8.5~kpc but more than 60\% of the corresponding 70$\mum$ flux, this is explained by the rapid decrease of infrared emission with radius. The corresponding values for 24 and 160$\mum$ are similar to a few percent at each radius. Tab.~\ref{tab.ancillary} summarizes the source and resolution of the ancillary data used.
\subsubsection{Surface brightness}
\begin{figure}
	[htbp]  
	\begin{flushleft}
		\includegraphics[angle=0,width=8.8cm]{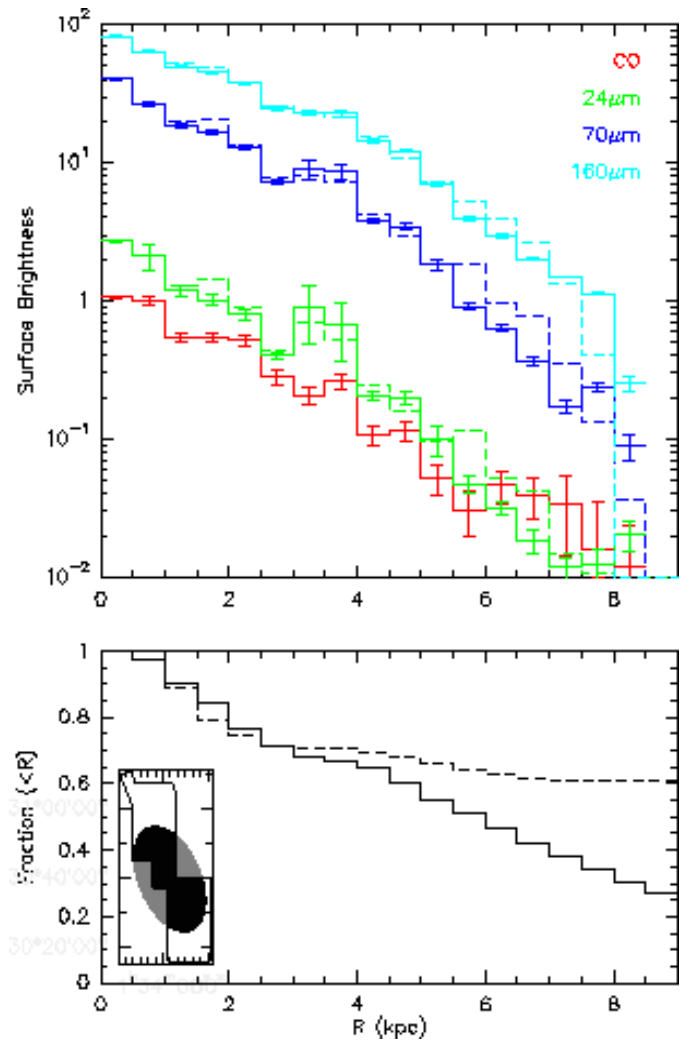} \caption{\label{fig.S_radius}(Color online) \emph{Top}: Radial distribution of the CO and IR surface brightness in units of MJy~sr$^{-1}$ for infrared data and K$\kms$ for CO. For the dust tracers, the surface brightness is computed both on the region covered by our CO observations (solid line) and on the whole elliptic annuli for each radius (dashed lines), showing that our CO sampled area is representative of the galaxy for each radius. \emph{Bottom}: Cumulative fractions of the area (solid line) and 70$\mum$ surface brightness (dashed line) in the region covered by our CO observations {  (black area in the lower left diagram)} compared to the whole elliptic region of radius less than R { (sum of grey and black areas at a given R).}} 
	\end{flushleft}
\end{figure}

The top part of Fig.~\ref{fig.S_radius} presents the radial variations in average surface brightness of the CO and infrared data from \emph{Spitzer} at 24$\mum$, 70$\mum$ and 160$\mum$, the solid lines correspond to averages taking into account only our CO mapped area and the dashed lines to the whole elliptical annuli for a given radius. For each tracer, the similarity between these two averages is an indication that the area mapped in CO is representative of the M33 disk in its entirety. The 4 datasets have similar exponential decreases with radius i.e. $F_{\nu}(r)\propto {\rm e}^{-r/L}$ with a corresponding scale length $L$. Table~\ref{tab.lenghtscales} summarizes the scale lengths computed by  least square fittingÊ  for CO and IR data, {  and Table~\ref{tab.lenghtscales_comp} scalelength values from surveys presented in Table~\ref{tab.surveys}.} The infrared data scale lengths are in agreement with the values in \citet{Tabatabaei.2007b} and \citet{Verley.2009}  albeit slightly smaller as the IR tracers seem to drop marginally more rapidly in our CO mapped region (solid lines in Fig.\ref{fig.S_radius}) compared to the whole galaxy (dashed lines in the same figure). The scale length of CO is higher than the ones of $24\mum$ and $70\mum$ and similar to the $160\mum$.
\begin{table}
	\begin{minipage}
		{88mm} \caption{\label{tab.ancillary}Ancillary data used in this paper.} 
		\begin{tabular*}
			{88mm}{@{\extracolsep{\fill}}lll} \hline\hline\noalign{\smallskip}  Wavelength & Resolution & Telescope\\
			\noalign{\smallskip}\hline\noalign{\smallskip}  160$\mum$ & $40\arcsec$ & \emph{Spitzer} MIPS \footnotemark[1]\\
			70$\mum$ & $18\arcsec$ & \emph{Spitzer} MIPS \footnotemark[1]\\
			24$\mum$ & $6\arcsec$ & \emph{Spitzer} MIPS \footnotemark[1]\\
			8$\mum$ & $2\arcsec$ & \emph{Spitzer} IRAC \footnotemark[2]\\
			6570 \AA\,(H$\alpha$)& $2\arcsec$ (pixel size) & KPNO \footnotemark[3]\\
			1500 \AA& $5\arcsec$ & GALEX FUV \footnotemark[4]\\
			\noalign{\smallskip}\hline 
		\end{tabular*}
		\footnotetext[1]{\citet{Hinz.2004,Tabatabaei.2007a}} \footnotetext[2]{\citet{Verley.2007}} \footnotetext[3]{\citet{Hoopes.2001,Greenawalt.1998}} \footnotetext[4]{\citet{Thilker.2005}, retrieved from GR5 public release of the MAST archive} 
	\end{minipage}
\end{table}
\begin{figure}
	[tbp] 
	\begin{flushleft}
		\includegraphics[angle=0,width=8.8cm]{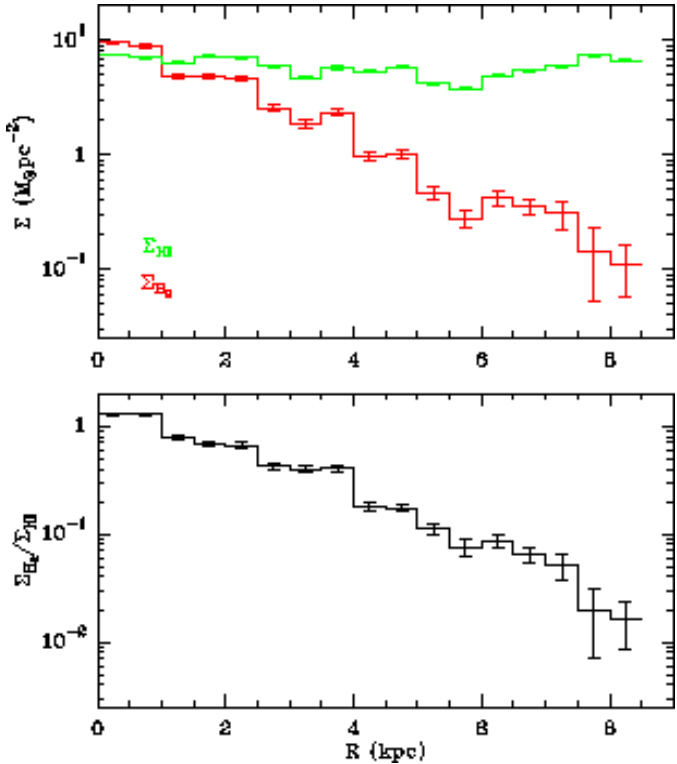} \caption{\label{fig.Sigma_radius} \emph{Top}: Radial distribution of the H$_{2}$ and \ion{H}{i} mass surface density. \emph{Bottom}: Radial distribution of the molecular to atomic mass fraction.} 
	\end{flushleft}
\end{figure}
\subsubsection{Mass surface densities}
We compute mass surface densities for atomic and molecular gas and study their radial distribution. For the molecular gas mass, we use the same hypotheses as in Sect~\ref{sec.H2mass} including a constant CO(2--1)/CO(1--0) factor. Taking a linearly decreasing value of this factor would modify only slightly the radial distribution, raising the most extreme outer points by a factor 1.5 and lowering the inner ones by a factor 1.1, at the price of added uncertainties that cannot be simply estimated. The top panel of Fig.~\ref{fig.Sigma_radius} shows the mass surface density of atomic and molecular gas as a function of radius. The atomic gas surface density is constant at an average value of ${\rm 6\pm2 ~M_{\sun}\,pc^{-2}}$, the molecular gas mass surface density follows the same trend as the CO surface brightness presented in Fig.~\ref{fig.S_radius}. The fraction of molecular to atomic gas in mass is shown in the bottom panel of Fig.~\ref{fig.Sigma_radius}. Given the constant atomic gas surface density, the fraction decreases following the molecular gas density from 1.2 to 0.015 from the center to the outer parts of the stellar disk. %A uniform increase in the $\ratio$ by a factor two to take into account the half solar metallicity of M33 would shift the curve a factor two upwards still retaining the trend of a steep fall of the molecular to atomic fraction with increasing radii.
\subsubsection{Star Formation Efficiency}
The star formation efficiency is usually defined as the ratio of the star formation rate over the mass of molecular gas available to form stars. 
$${\rm SFE=SFR/M_{H_{2}}}$$
{  Defined in this way, the SFE is the inverse of the molecular gas depletion time.}
Estimates from a variety of observations \citep{Blitz.2006,Kennicutt.1998,Hippelein.2003} give a value of the SFR which varies from 0.3 to 0.7${\rm ~M_{\sun}yr^{-1}}$ for the entire disk. A recent study by \citet{Verley.2009} using extinction corrected H$\alpha$, FUV, and infrared tracers gives a value of {  ${\rm 0.45 \pm 0.1~M_{\sun}yr^{-1}}$} for total star formation rate. This is our adopted value. 

\begin{figure}
	[tbp] 
	\begin{flushleft}
		\includegraphics[angle=0,width=8.8cm]{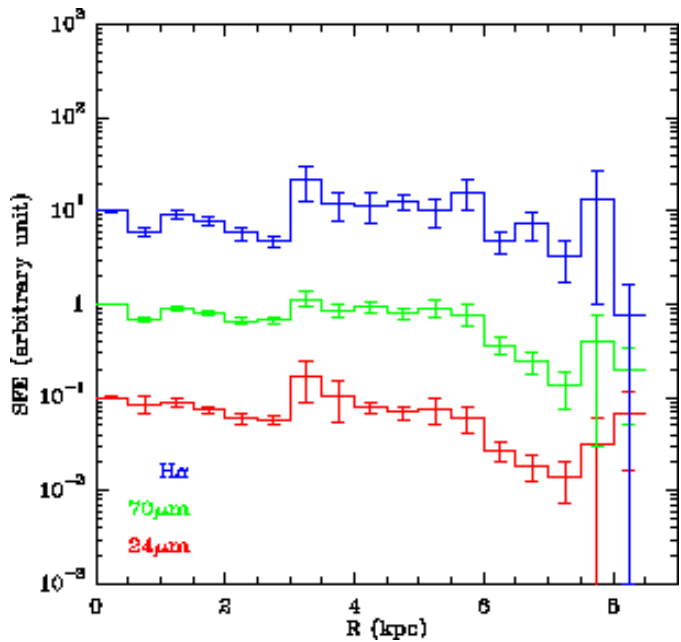} \caption{\label{fig.SFE_radius} Radial distribution of the Star Formation Efficiency in arbitrary units, all of the curves are normalized to the central value and shifted by 1 decade along the vertical axis.} 
	\end{flushleft}
\end{figure}

{  Our CO map does not cover the whole disk but covers respectively 63\%, 61\% and 59\% of the IR emission at 24$\mum$, 70$\mum$ and 160$\mum$ (see bottom panel of Fig.~\ref{fig.S_radius}) so the
SFR is $\sim  0.27$~M$_{\sun}$ yr$^{-1}$  within the area covered by our CO map.  With a $\ratio$ value 
of $4\times10^{20}\Xunit$ and an average 2--1/1--0 line ratio of 0.73 (see Sect.~\ref{sec.H2mass} for details), the molecular 
gas mass within the CO map is $1.9\times10^8$~M$_{\sun}$.  With these values, the SFE is ${\rm 1.4\times10^{-9}~yr^{-1}}$ or a molecular gas depletion time ${\rm 7\times10^{8}~yr}$.
An SFE of ${\rm 1.6\times10^{-9}~yr^{-1}}$ is a factor $\sim 3$ higher than that found in large spiral galaxies \citep[sample in,][]{Kennicutt.1998,Murgia.2002,Leroy.2008}  
Earlier studies used lower $\ratio$ values and thus obtained even higher SFEs (or 
shorter H$_2$ depletion times) as is shown in Table~\ref{tab.avec_comp}.

In order to bring the SFE of M33 to the same level as in large spirals, the $\ratio$
factor would have to be about 3 times the value we use whereas previous studies used even smaller values \citep{Wilson.1990,Corbelli.2003,Heyer.2004,Engargiola.2003,Rosolowsky.2007a,Gardan.2007}. Through Virial mass estimates and 
$^{13/12}$CO line ratios, Braine et al (submitted to A\&A) support a value of about 
$\ratio=4\times10^{20}~\Xunit$ in the disk of M33, possibly closer to the Galactic value in the inner disk.

Figure~\ref{fig.SFE_radius} presents the radial distributions of the SFE as calculated from the H$\alpha$, 
24, and 70$\mum$ emission. }
%In the region of radii smaller than R$_{25}$, our map accounts for respectively 63\%, 61\% and 59\% of the IR emission at 24$\mum$, 70$\mum$ and 160$\mum$ (see bottom panel of Fig.~\ref{fig.S_radius}) we therefore adopt a star formation rate for our CO mapped region of ${\rm 0.3~M_{\sun}yr^{-1}}$. Using the same prescription as in the previous paragraph, we find a molecular gas mass over our CO mapped area of ${\rm 1.9\times10^8~M_{\sun}}$. The SFE for our CO mapped area is then ${\rm 1.6\times10^{-9}~yr^{-1}}$ or a molecular gas depletion time ${\rm 6\times10^{8}~yr}$. %in agreement with the molecular gas depletion time value of ${\rm 2-5\times10^{8}~yr}$ found by \citet{Gardan.2007} in a less sensitive subset of our CO map. 
%This SFE value is a factor 2-4 times higher than that found in large spiral galaxies \citep[sample in,][]{Kennicutt.1998,Murgia.2002}. One caveat is that the $\ratioo$ ratio might be even higher which would raise the H$_{2}$ mass.
% Figure~\ref{fig.SFE_radius} presents the radial distributions of the SFE for M33. 
The units are arbitrary as we are interested in showing the radial trend. The SFR were taken as being directly proportional to the H$\alpha$ and IR luminosities. H$\alpha$ and FIR emission tend to show opposite biases -- the H$\alpha$ suffers extinction where the FIR is strong and the FIR is weak where little dust is present but the H$\alpha$ is unaffected. If one were to compute the star formation efficiency taking into account the total gas mass, the SFE would drop dramatically with increasing radius as opposed to the SFE computed using only molecular gas that does not appear to vary with radius. {  Table~\ref{tab.avec_comp} summarizes the masses, star formation rates and depletion times for this study and previous surveys of M33.}
\subsection{\label{sec.SFtracers} The ISM and star formation tracers} 
\begin{figure}
	[tbp] 
	\begin{flushleft}
		\includegraphics[angle=0,width=8.8cm]{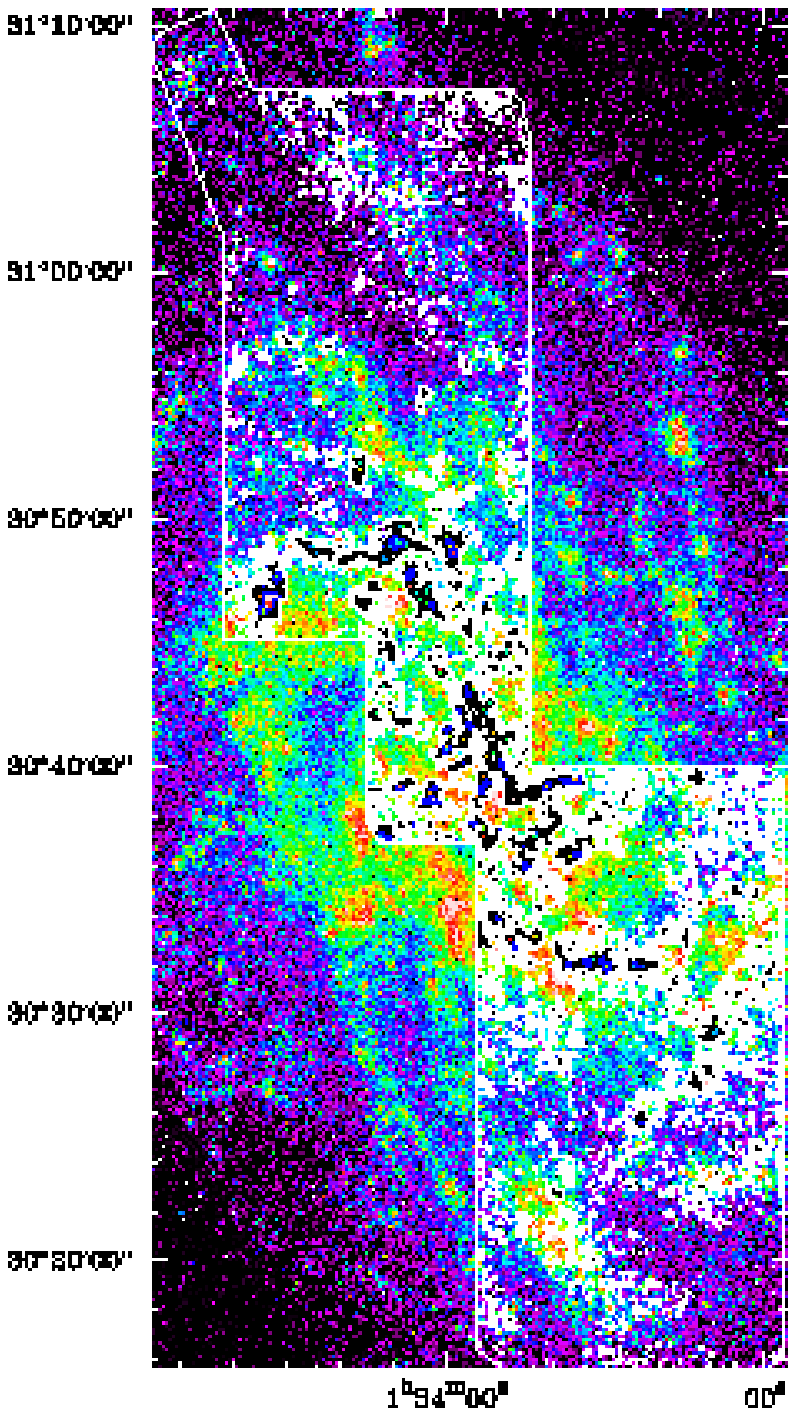} \caption{\label{fig.FUV_CO}Galex FUV image with \ICO main beam contours of 1~(white), 2,~4~(black), 8~(blue) K$\kms$. The beam size is shown as a white dot in the lower left corner.
 Blow ups are available in the online version.} 
	\end{flushleft}
\end{figure}

In this section, we compare the CO emission with tracers of star formation. We present a series of figures (\ref{fig.FUV_CO} to~\ref{fig.70mu_CO}) comparing the CO emission, as contours, overlaid on images of the 8, 24 and 70$\mum$ FIR emission as well as the H$\alpha$ and FUV emission (\emph{Blow ups of these 5 figures are available in the online version of this article}). The 160$\mum$ image is at lower resolution and brings little morphological information not present in the 70$\mum$ image.
\begin{table}
	\begin{minipage}
		{88mm} \caption{\label{tab.lenghtscales}Exponential scale length $L$ in kpc for Spitzer MIPS and CO emission computed over the area mapped in CO.
		% Fitted curves are dominated by the brightest part of the disk if we fit a 
		} 
		\begin{tabular*}
			{88mm}{@{\extracolsep{\fill}}lrrr} 
			\hline\hline\noalign{\smallskip} 
			 & [0.5-3.5]~kpc & [3.5-7]~kpc  & [0.5-7]~kpc \\
			\noalign{\smallskip}\hline\noalign{\smallskip}
			24$\mum$&$1.36\pm0.05$&$1.02\pm0.07$&$1.40\pm0.03$ \\
			70$\mum$&$1.51\pm0.07$&$1.05\pm0.06$&$1.48\pm0.04$ \\
			160$\mum$&$2.26\pm0.12$&$1.30\pm0.06$&$1.83\pm0.07$ \\
			CO & $2.0\pm0.2$&$1.6\pm0.3$&$1.9\pm0.1$ \\
			\noalign{\smallskip}\hline
		\end{tabular*}
	\end{minipage}
\end{table}
\begin{table}
	\begin{minipage}
		{88mm} \caption{\label{tab.lenghtscales_comp}Comparison with  previous surveys of CO emission exponential scale lengths $L$ in kpc.
		}
		\begin{tabular*}
			{88mm}{@{\extracolsep{\fill}}lrr} 
			\hline\hline\noalign{\smallskip} 
			 Study & Range&  Scalelength \\
			       &  (kpc) &(kpc)\\
			\noalign{\smallskip}\hline\noalign{\smallskip} 
			This paper & 0--7 &$1.9\pm0.1$\\
			\citet{Gardan.2007}&2--6&$1.4\pm0.1$\\
			\citet{Engargiola.2003} & 0--7& $1.4\pm0.1$\\
			\citet{Corbelli.2003}& 0--6& 2.5\\
			\noalign{\smallskip}\hline
		\end{tabular*}
	\end{minipage}
\end{table}

\begin{figure}
	[tbp] 
	\begin{flushleft}
		\includegraphics[angle=0,width=8.8cm]{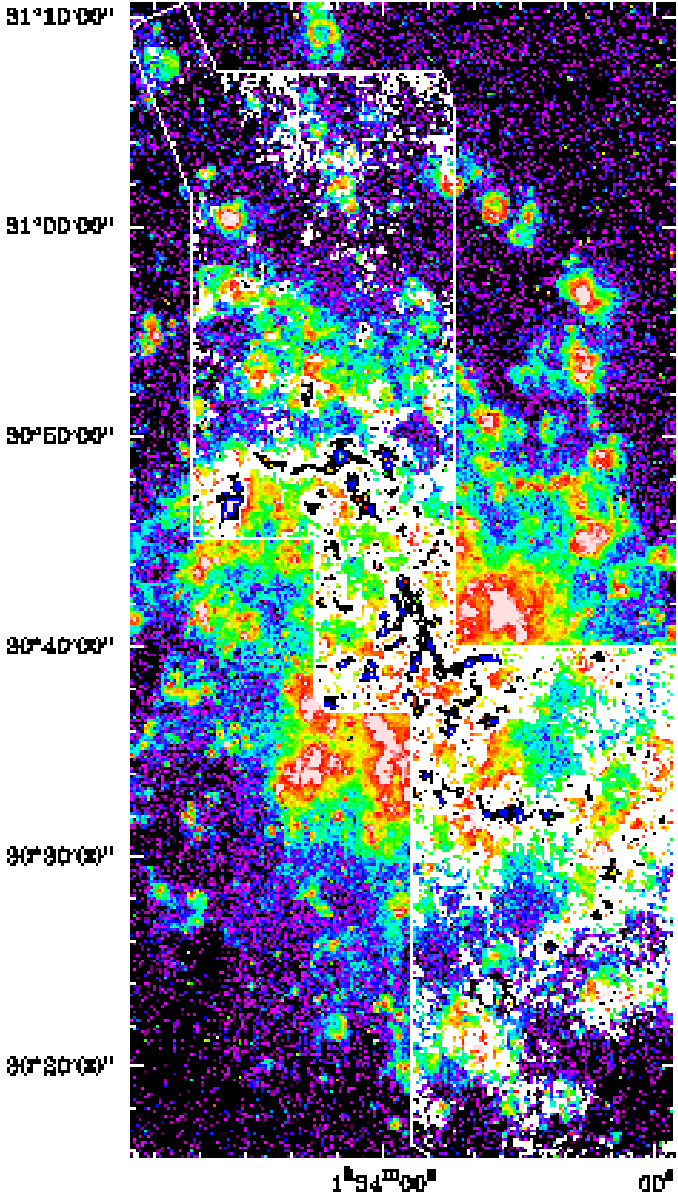} \caption{KPNO H${\alpha}$ image with \ICO main beam contours of 1~(white), 2,~4~(black), 8~(blue) K$\kms$. The beam size is shown as a white dot in the lower left corner.
 Blow ups are available in the online version.} 
	\end{flushleft}
\end{figure}
\begin{figure}
	[tbp] 
	\begin{flushleft}
		\includegraphics[angle=0,width=8.8cm]{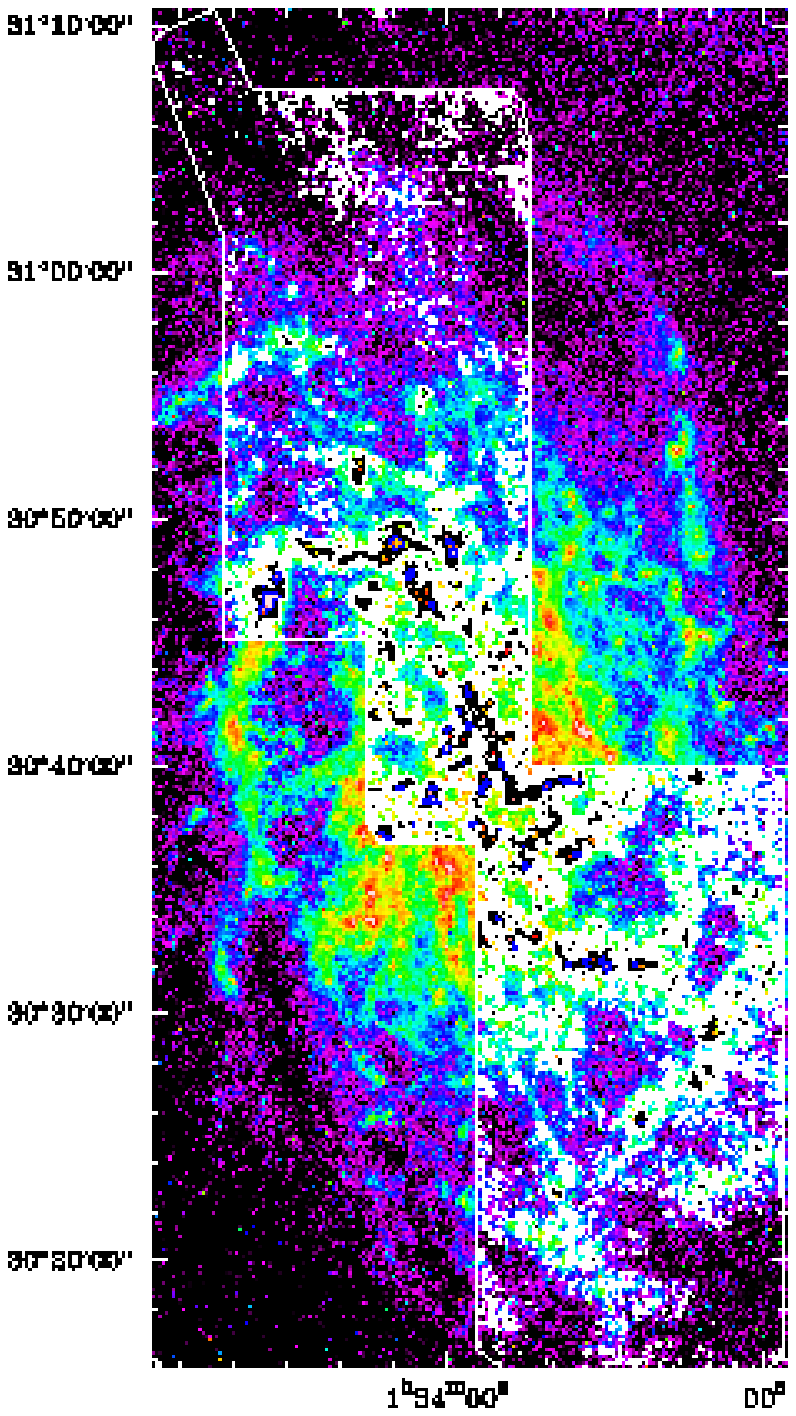} \caption{Spitzer 8$\mum$ image with \ICO main beam contours of 1~(white), 2,~4~(black), 8~(blue) K$\kms$. The beam size is shown as a white dot in the lower left corner.
 Blow ups are available in the online version.} 
	\end{flushleft}
\end{figure}
\begin{figure}
	[tbp] 
	\begin{flushleft}
		\includegraphics[angle=0,width=8.8cm]{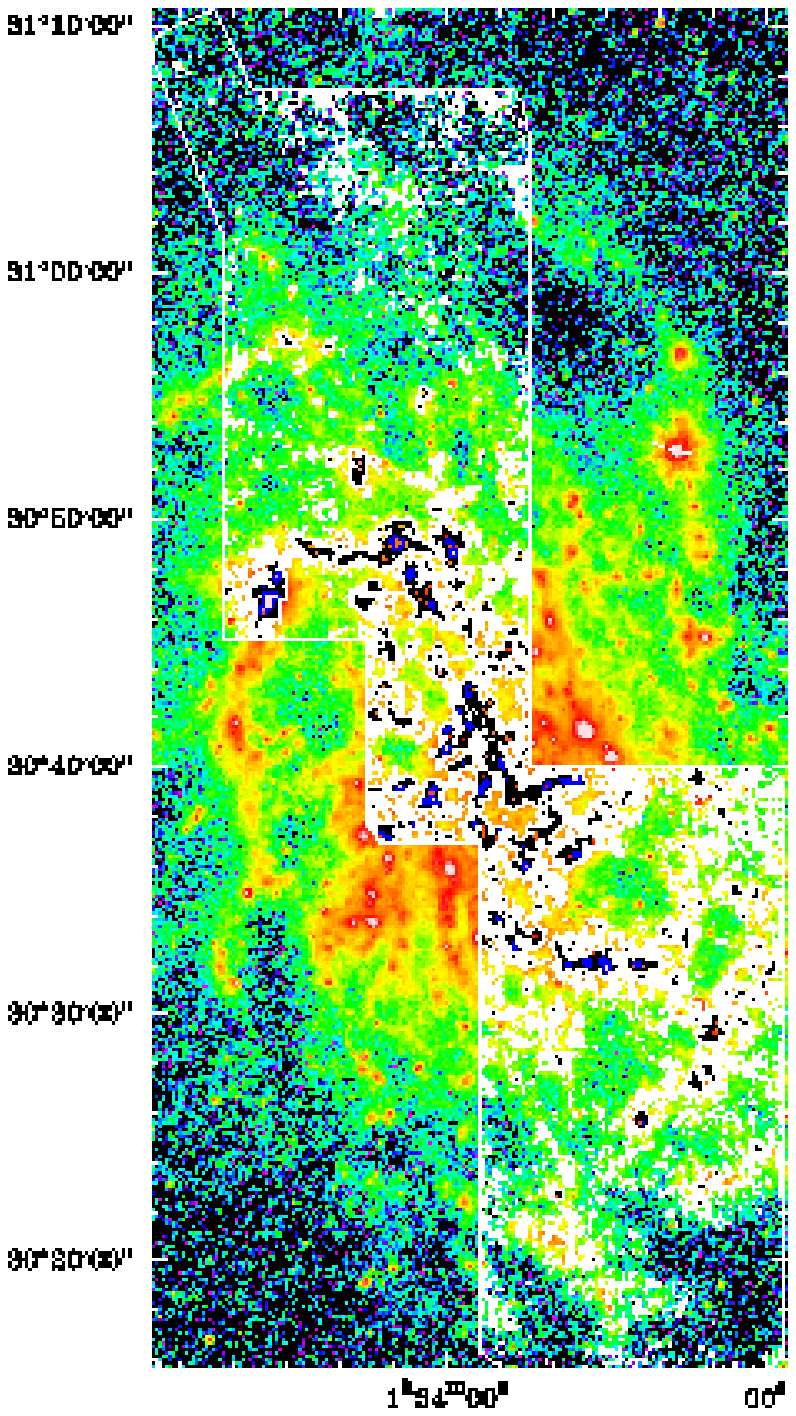} \caption{Spitzer 24$\mum$ image with \ICO main beam contours of 1~(white), 2,~4~(black), 8~(blue) K$\kms$. The beam size is shown as a white dot in the lower left corner.
 Blow ups are available in the online version.} 
	\end{flushleft}
\end{figure}
\begin{figure}
	[tbp] 
	\begin{flushleft}
		\includegraphics[angle=0,width=8.8cm]{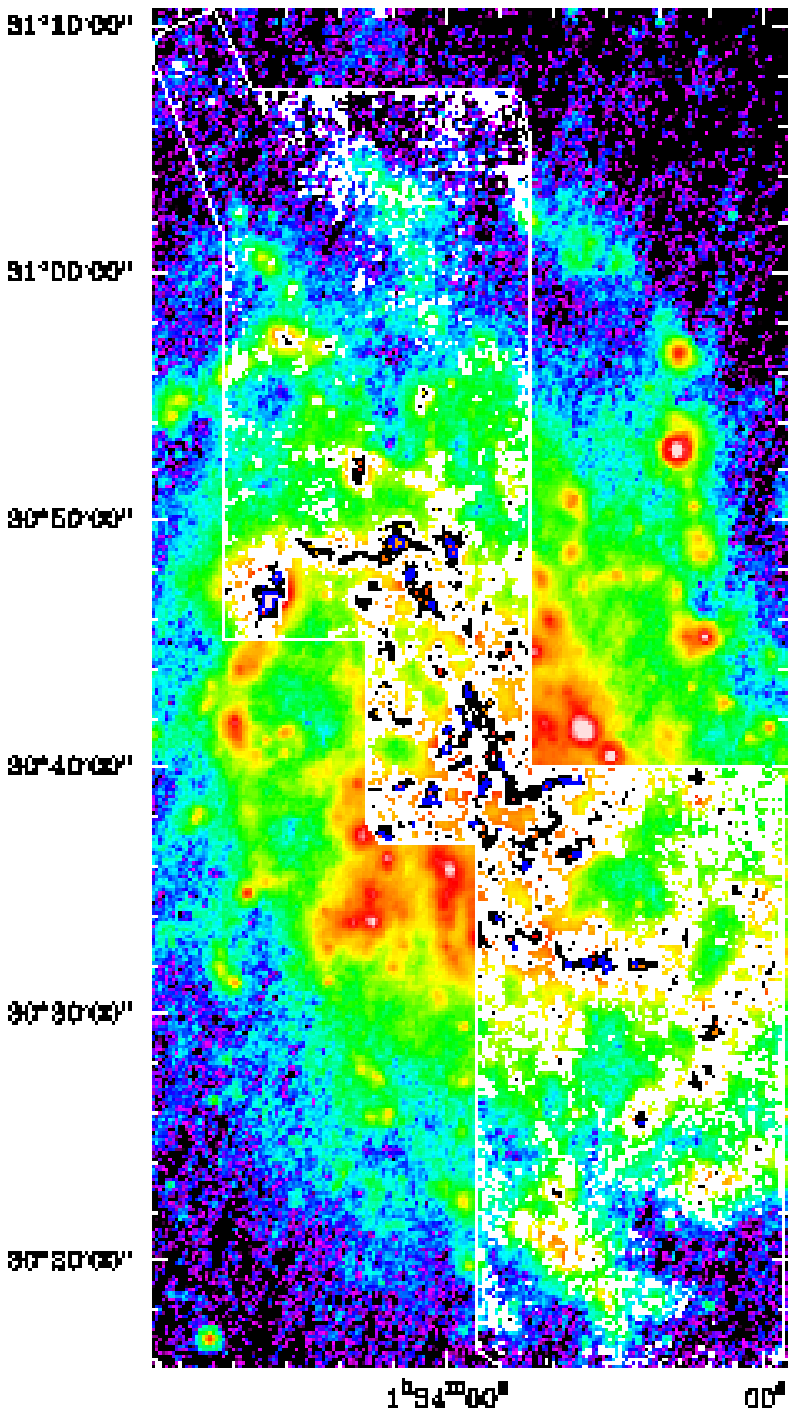} \caption{\label{fig.70mu_CO}Spitzer 70$\mum$ image with \ICO main beam contours of 1~(white), 2,~4~(black), 8~(blue) K$\kms$. The beam size is shown as a white dot in the lower left corner.
 Blow ups are available in the online version.} 
	\end{flushleft}
\end{figure}
\begin{figure}
	[tbp] 
	\begin{flushleft}
		\includegraphics[angle=0,width=8.8cm]{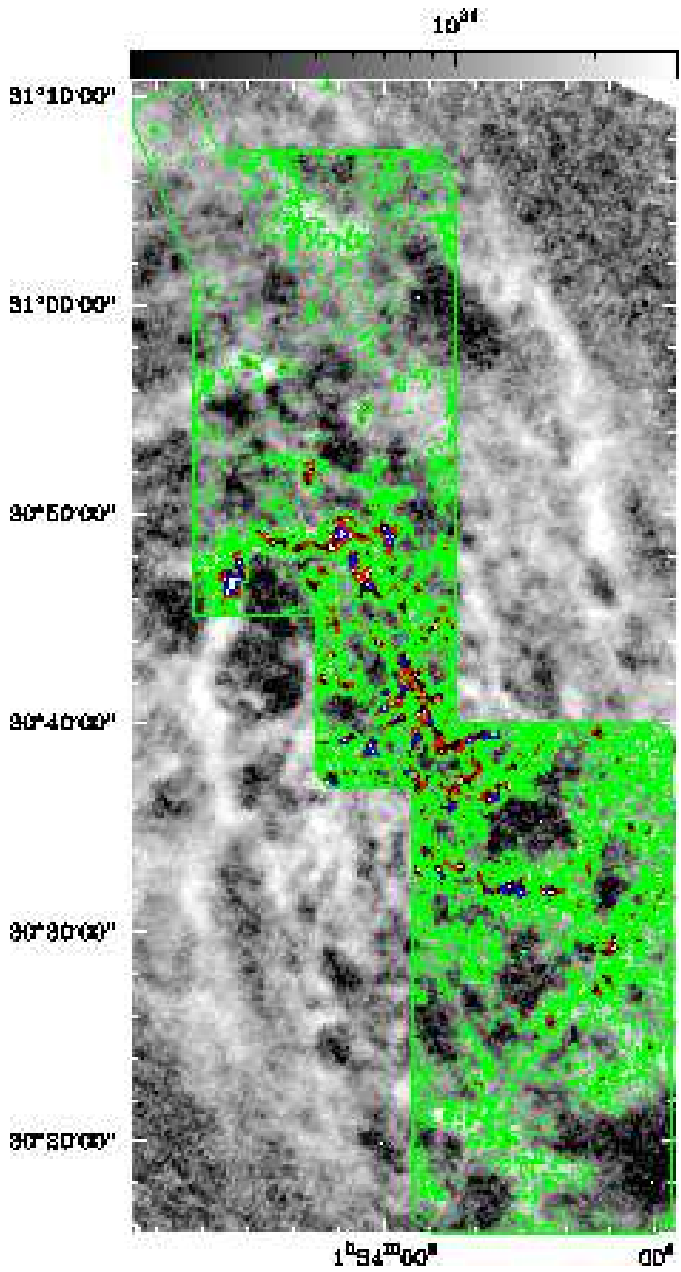} \caption{\label{fig.HI_CO}VLA \ion{H}{i} zeroth moment image with \ICO main beam contours of 1~(green), 2~(red), ~4(black), 8~(blue) K$\kms$. The beam size is shown as a white dot in the lower left corner.
 Note how the CO emission follows the bright \ion{H}{i} and outlines the \ion{H}{i} holes. Blow ups are available in the online version.} 
	\end{flushleft}
\end{figure}
\begin{figure}
	[tbp] 
	\begin{flushleft}
		\includegraphics[angle=270,width=8.8cm]{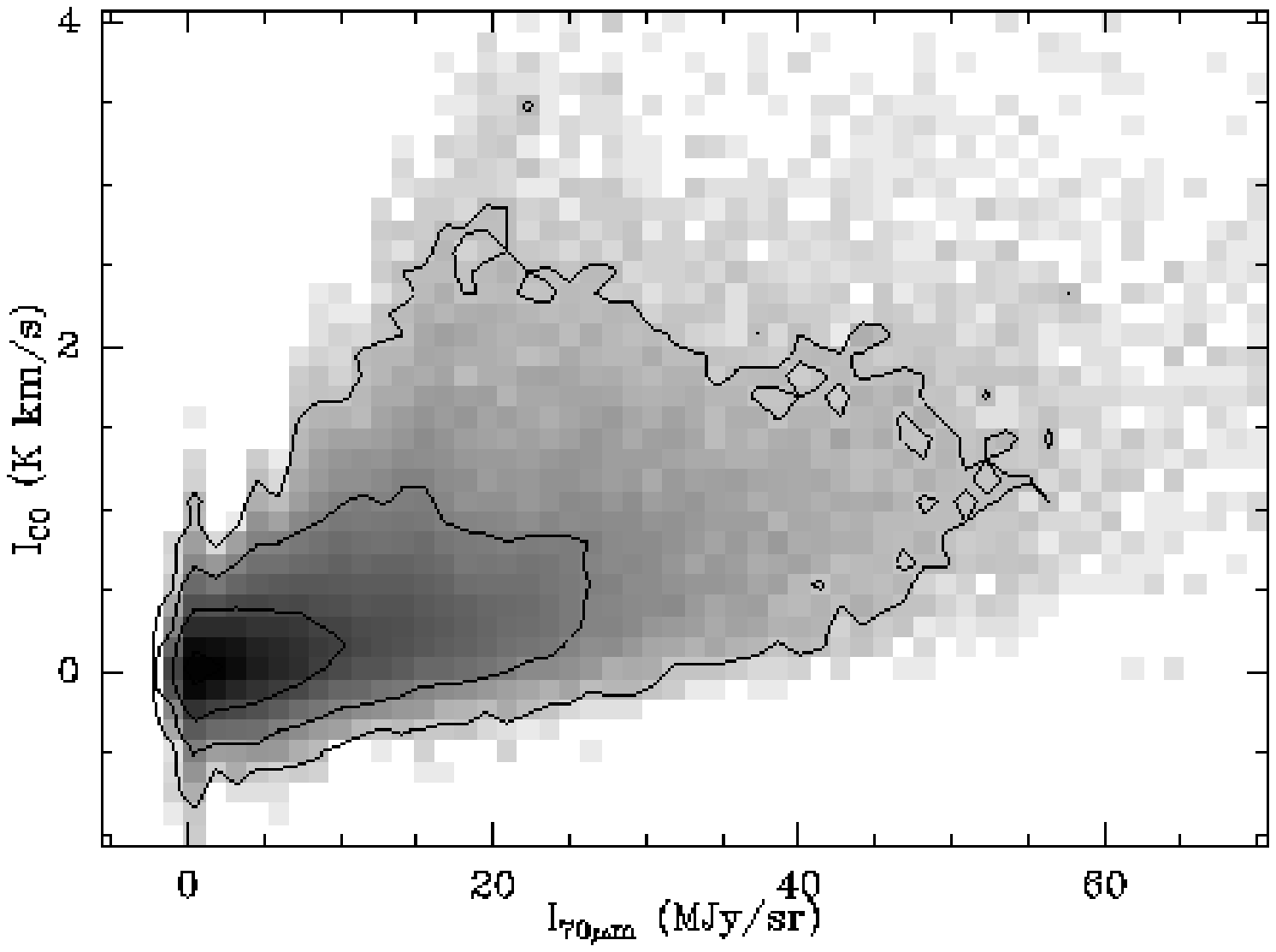} \caption{\label{fig.cross_histo70}Scatter plot between CO integrated intensity and 70$\mum$ emission.} 
	\end{flushleft}
\end{figure}
%\begin{figure}
%	[tbp] 
%	\begin{flushleft}
%		\includegraphics[angle=270,width=8.8cm]{figures/lores_cross_histoHI.eps} \caption{\label{fig.cross_histoHI}{  Scatter} plot between CO and \ion{H}{i} integrated intensity.} 
%	\end{flushleft}
%\end{figure}

While the dynamic range in the CO image is much lower than in the images of star formation tracers, the correspondence is globally excellent.  Figure~\ref{fig.cross_histo70} shows the scatter plot between the CO integrated intensity and 70$\mum$ emission. 

\begin{figure}
	[tbp]  
	\begin{flushleft}
		\includegraphics[angle=0,width=8.8cm]{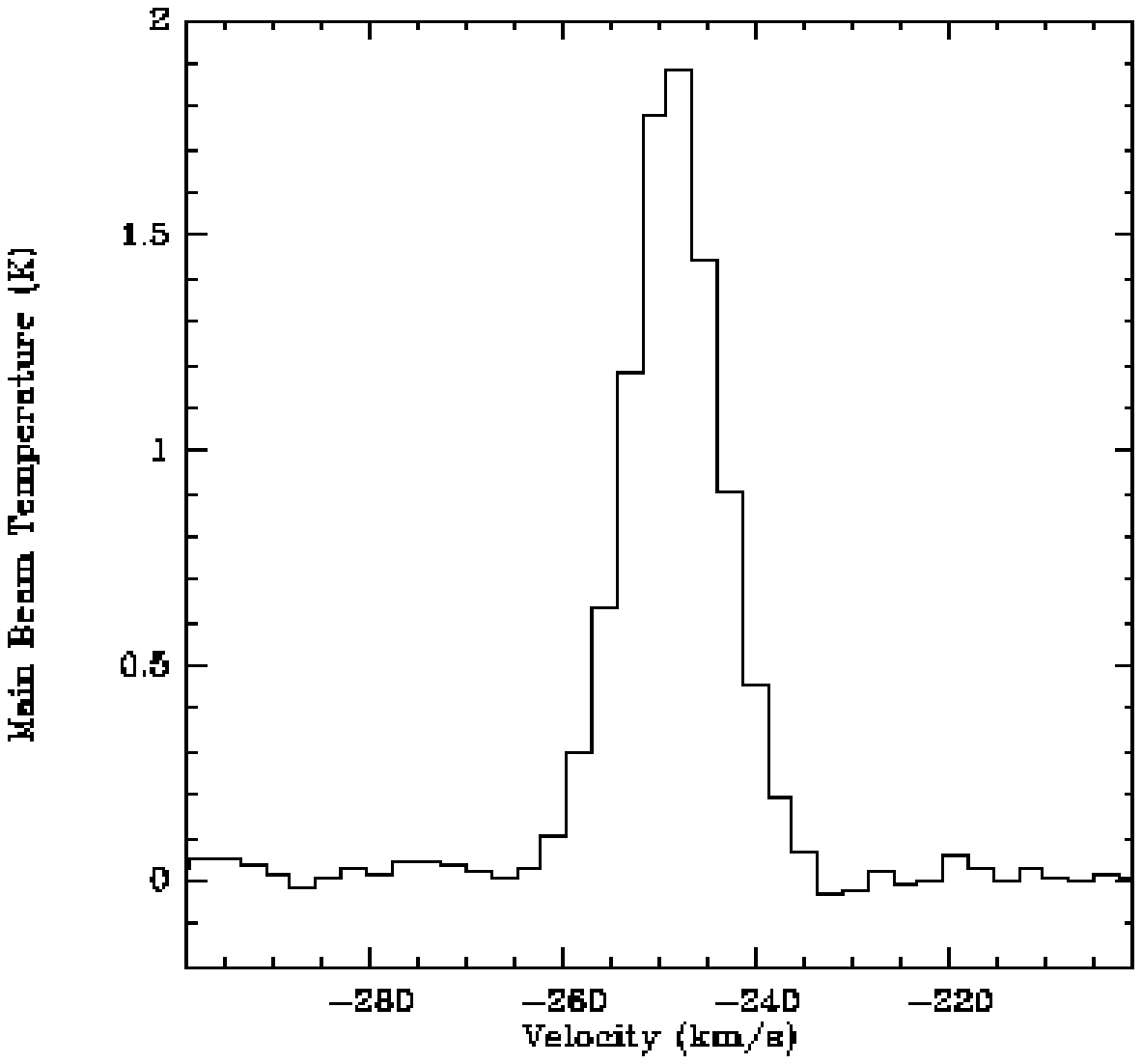} \caption{\label{fig.spec_max}Spectrum of a bright cloud at position ({$\alpha=1^\mathrm{h}34^\mathrm{m}09\fs4$}, {$\delta=+30\degr49\arcmin06\arcsec$}, J2000), unlike other strong peak of CO integrated intensity, it lacks strong emission in star formation tracers (see Sect.Ê\ref{sec.SFtracers})} 
	\end{flushleft}
\end{figure}

Figures~\ref{fig.FUV_CO}--\ref{fig.70mu_CO} show that in almost all cases the CO emission is found where star formation is detected and in general the CO emission is stronger as the level of SF increases. The weaker CO contours outline the regions with a remarkable precision (see online figures 1 to 12 for the blowups). A partial exception is found near ($\alpha=1^\mathrm{h}34^\mathrm{m}09\fs4$, $\delta=+30\degr49\arcmin06\arcsec$, J2000) where the CO emission is extremely strong while the SF tracers show rather weak emission. The CO spectrum in the direction of that particular cloud is shown in Fig.~\ref{fig.spec_max}.

A true exception {  to the FIR--HI--CO correlation} is the interarm cloud at ($\alpha=1^\mathrm{h}34^\mathrm{m}16\fs7$, $\delta=+30\degr59\arcmin3\arcsec$, J2000) dubbed ``Lonely Cloud'' by \citet{Gardan.2007}. It is found in a region with very little if any star formation and relatively weak \ion{H}{i} emission, despite fairly strong CO (and even $^{13}$CO) emission. When \ion{H}{i} brightness thresholds higher than 10~K are used to create the masked zero-moment CO maps, the Lonely Cloud disappears. While this cloud was discovered in the first set of observations, it remains a unique object as the observations of the central and southern regions, despite considerably lower noise levels than in \citet{Gardan.2007}, have not revealed similar molecular clouds. 

In the outer parts, the FIR emission is typically quite inconspicuous, even where CO has been detected, but the H$\alpha$ emission is often quite strong. In the central regions, the alignment of the emission at the different wavelengths is nearly perfect, especially between CO and FIR. Further out, however, notable offsets become more common although there is a clear association between the CO and FIR/H$\alpha$ emitting regions. The offsets are less than 100~pc, particularly in comparison with the FIR. 

The CO emission follows HI-defined arm-like structures very closely, completely avoiding the \ion{H}{i} holes. As can be seen in Fig.~\ref{fig.HI_CO} and corresponding blow ups in the online version of the paper, the overall correlation between \ion{H}{i} and CO emission is very good. CO is generally found in HI-rich regions although not all of the \ion{H}{i} peaks show CO emission and not all CO emission falls on \ion{H}{i}.
% peaks but none of the \ion{H}{i} holes shows detectable CO emission.
  It is not clear what the source of the 70$\mum$ emission is in for example the \ion{H}{i} (and CO) hole around ($\alpha=1^\mathrm{h}34^\mathrm{m}25$, $\delta=+30\degr35\arcmin00\arcsec$, J2000) as the amount of neutral gas is very low. The emission in this region is lower in the 24 and 8$\mum$ wavebands, so perhaps the 70 micron emission appears stronger than in reality due to the poorer spatial resolution. The upcoming Herschel {\tt HERM33ES} observations should answer this.

%NIR also to show how weak the stellar emission is ?
%Interesting spots for zooms?
%outer cloud
%double peak 
%\clearpage

%\section{The arm-interarm contrast in the ISM of M33}
%\citet{Regan.1994} show that two stellar spiral arms can be identified .  
%They cover essentially the inner disk where the stellar surface density is higher than that of the gas.
%Fig XX shows the regions we have considered as being spiral arms -- these are roughly the same as in \citet{Rosolowsky.2007}. 
%To evaluate the arm-interarm contrast, we have masked all but the arm regions and calculated the average \ion{H}{i} and CO brightnesses. then we did the same for the interarm regions (i.e. the rest).  For the HI, since the whole galaxy is available, we did this for the regions observed in CO and for M33 as a whole.

\subsection{Examples of interesting spectra} 
\begin{figure}
	[tbp]  
	\begin{flushleft}
		\includegraphics[angle=0,width=8.8cm]{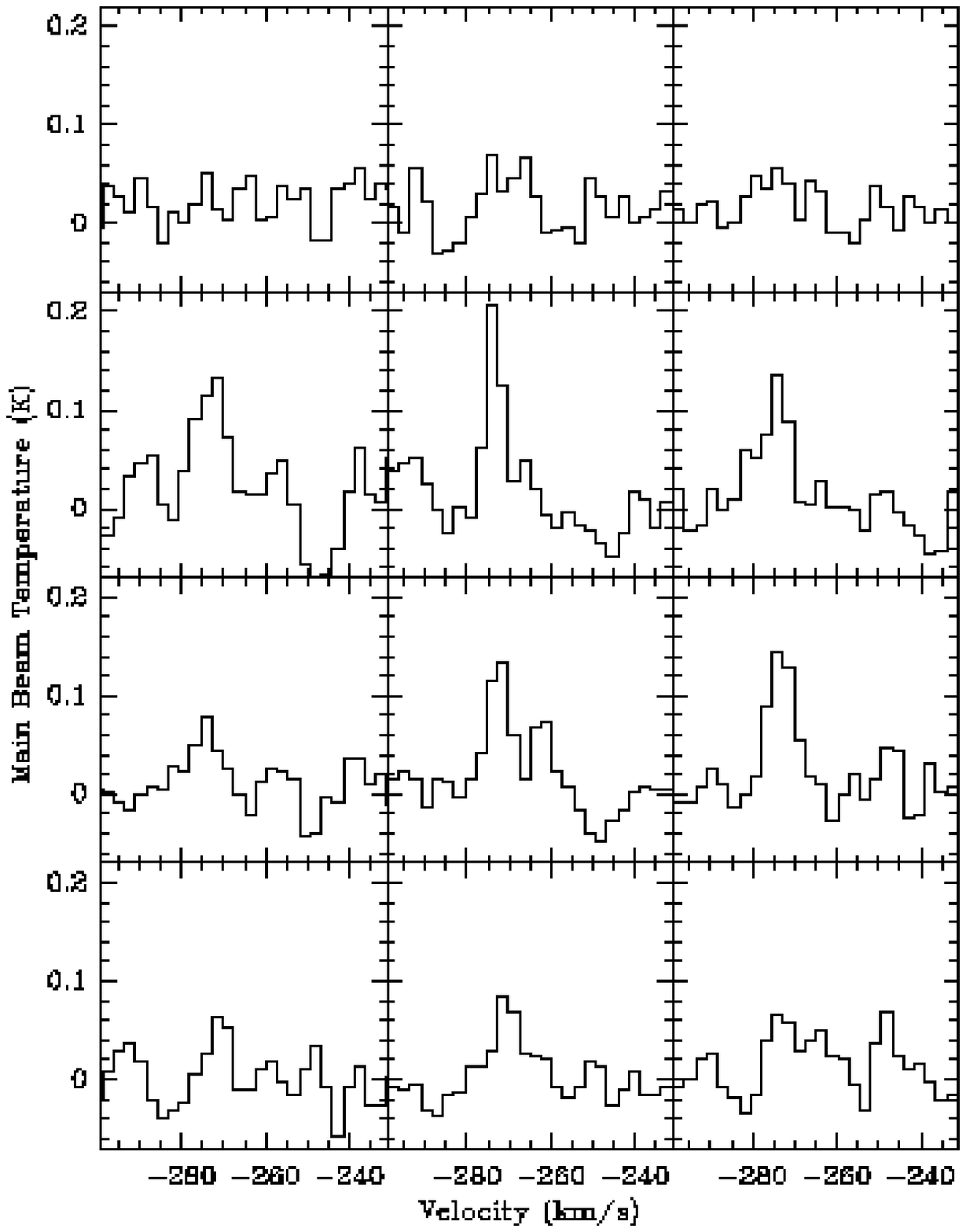} \caption{\label{fig.faroutCO_spec}Spectra of the outer disk cloud ({$\alpha=1^\mathrm{h}34^\mathrm{m}50\fs8$}, {$\delta=+31\degr08\arcmin26\arcsec$}, J2000) shown with $6\arcsec$ spacing between panels. If the secondary peak at -266$\kms$, seen in the second spectrum from the bottom in the middle column, is real then the cloud is small as it is only seen within a single beam.} 
	\end{flushleft}
\end{figure}

\begin{figure}
	[tbp]  
	\begin{flushleft}
		\includegraphics[angle=0,width=8.8cm]{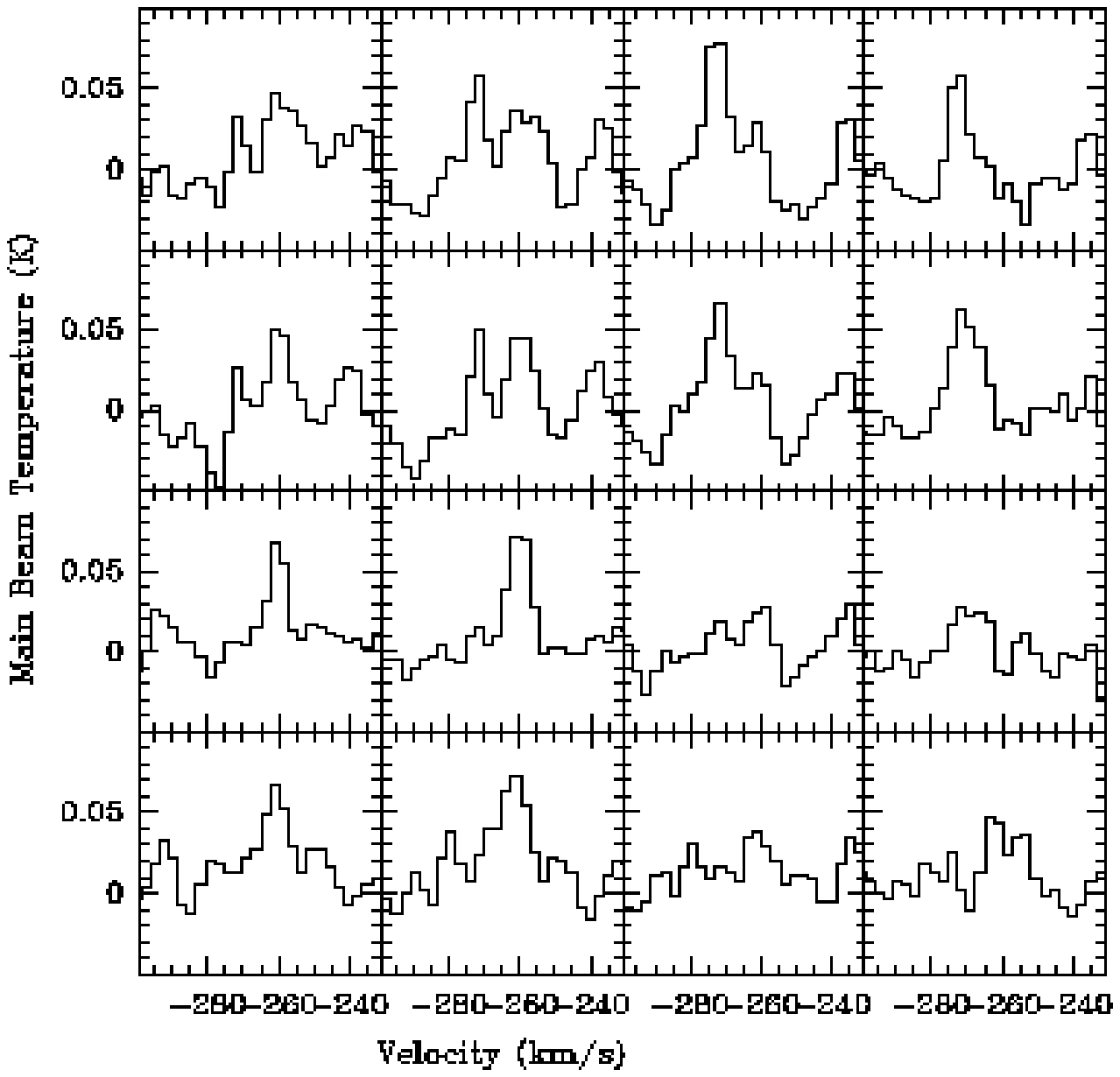} \caption{\label{fig.faroutCO_spec2}Spectra of two more outer disk clouds at different velocities. The lower left panel is at ({$\alpha=1^\mathrm{h}34^\mathrm{m}21\fs7$}, {$\delta=+31\degr04\arcmin13\arcsec$}, J2000) and the panel spacing is $6\arcsec$. } 
	\end{flushleft}
\end{figure}

The spectra shown in Figures~\ref{fig.spec_max},~\ref{fig.faroutCO_spec} and~\ref{fig.faroutCO_spec2} are just a few among a huge number but are intended to show ``special'' positions. Fig~\ref{fig.spec_max} shows a very strong CO profile towards a region without strong star formation \citep[note good agreement with ][]{Engargiola.2003}. The line is very strong with a line width of about 11$\kms$ at half-intensity. The other spectra shown are interesting because the CO lines are detected very far from the center of M33, despite the low metallicity and surface mass density. Fig.~\ref{fig.faroutCO_spec} shows the second molecular cloud detected at or beyond R$_{25}$ in M33. A secondary peak offset in velocity by about 8$\kms$ may be present. Fig~\ref{fig.faroutCO_spec2} shows further spectra near R$_{25}$ which clearly is composed of emission from two clouds, the cloud to the SE (of the figure) at -260$\kms$ and the one to the NW at about -273$\kms$. The spectral resolution of the data is barely sufficient to measure the linewidths of the clouds, which are typically less than 5$\kms$ at half power. This was also true of the spectra of distant clouds shown in \citet{Braine.2010} --- they are much narrower than in the H$_2$-rich inner disk. It is not clear whether the broad linewidths in the inner disk are because the GMCs themselves are much more massive or due to a superposition of clouds (or a single GMC in a region with significant diffuse emission).

The spectra shown also illustrate the dynamic range of the data acquired, roughly 100 in integrated intensity.
\section{Conclusions}
This work presents high-resolution maps of the atomic and molecular gas in 
the disk of M33 via observations of the \ion{H}{i}~21~cm and \mbox{CO(2--1)} lines.  
The whole disk out to 8.5 kpc is 
      covered in the \ion{H}{i} line and about 60\% of the emission in CO.  

Assuming the $\ratio$ factor to be twice that of the Galaxy, because of the sub-solar metallicity of M33, 
we estimate a molecular gas mass of $3.3\times10^8~{\rm M_{\sun}}$ roughly 20\% of the $1.4\times10^9~{\rm M_{\sun}}$ detected in the inner 8.5~kpc in \ion{H}{i}.
Azimuthally averaging, the \ion{H}{i} surface density is close to constant with radius but the H$_2$ decreases 
exponentially with a scale length of 1.9kpc.  The H$_2$/\ion{H}{i} mass ratio decreases from about unity to 1\%.

The correspondence between the peaks and holes in the distributions of molecular and atomic gas is excellent and
follows the peaks and troughs in the FIR, MIR, and H$\alpha$ images.  The SFE is approximately constant with radius,
suggesting that molecular gas is transformed into stars at a similar rate (assuming a similar IMF) at all galactocentric radii.
However, the SFE in the small, gas-rich, low-metallicity, blue spiral M33 appears {2-4 times} higher than what is observed 
in large nearby spirals.

The sensitivity of the survey is such that CO emission is detected far out in the disk of M33 although few 
clouds are present and the lines are much weaker in intensity.

%Lonely cloud still rather unique\\
%Excellent morphological correspondance between CO and the tracers of SF

\begin{acknowledgements}
We thank the IRAM staff in Granada for their help with the observations.
\end{acknowledgements}
\clearpage
\bibliographystyle{aa} 
\bibliography{/Users/gratier/Documents/Work/These/Biblio/biblio}
%\bibliography{biblio}

%\end{document}
\begin{appendix}
\section{Online Figures}
{ 
\begin{figure*}
	[tbp] 
	\begin{flushleft}
		
		\includegraphics[angle=0,width=18cm]{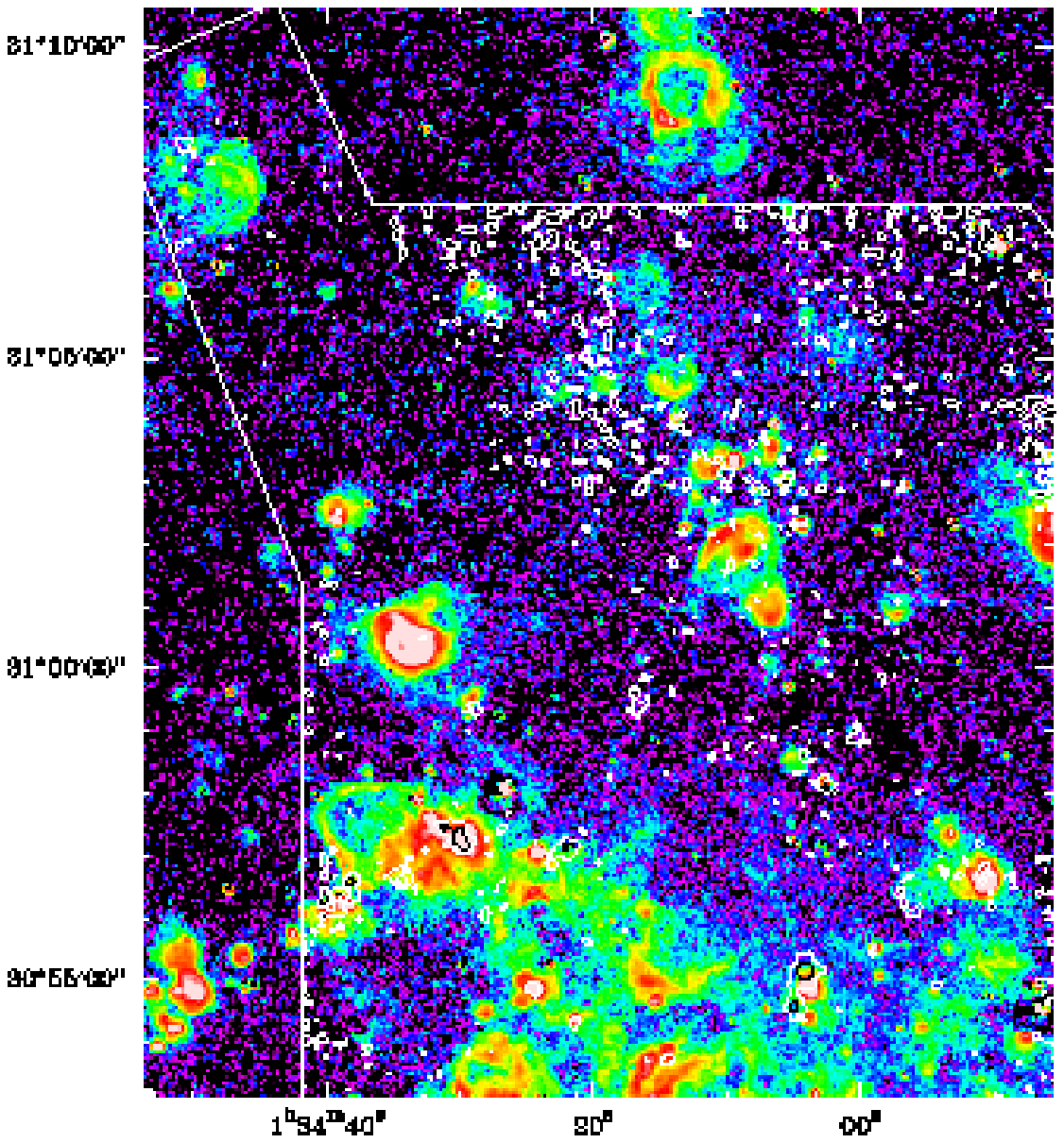}
		\caption{Northern part of KPNO H${\alpha}$ image with \ICO main beam contours of 1~(white), 2,~4~(black), 8~(blue) K$\kms$. The beam size is shown as a white dot in the lower left corner.
} 
	\end{flushleft}
\end{figure*}
}

{ 
\begin{figure*}
	[tbp] 
	\begin{flushleft}
		
		\includegraphics[angle=0,width=18cm]{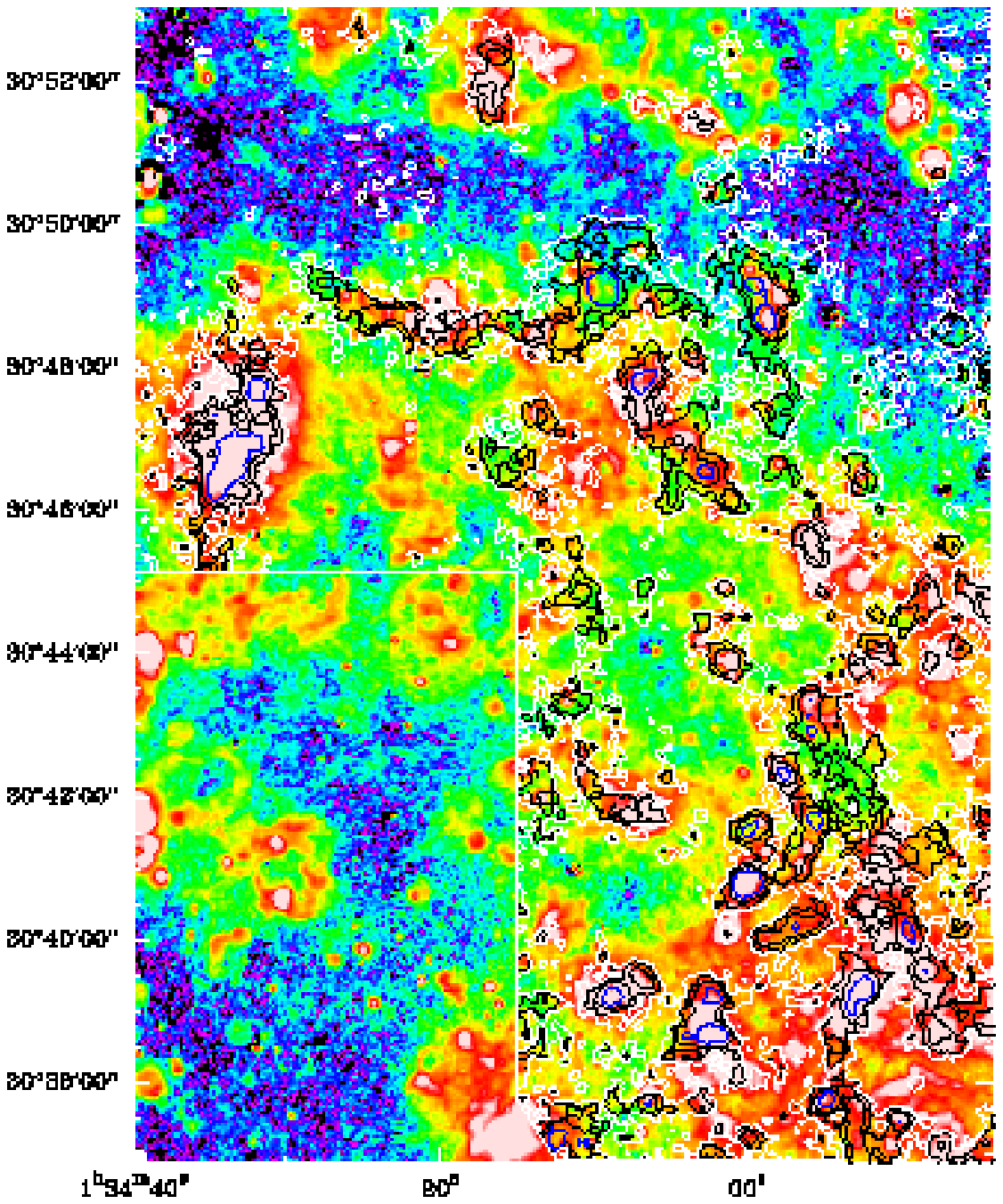}
		\caption{Center part of KPNO H${\alpha}$ image with \ICO main beam contours of 1~(white), 2,~4~(black), 8~(blue) K$\kms$. The beam size is shown as a white dot in the lower left corner.
} 
	\end{flushleft}
\end{figure*}
}

{ 
\begin{figure*}
	[tbp] 
	\begin{flushleft}
		
		\includegraphics[angle=0,width=18cm]{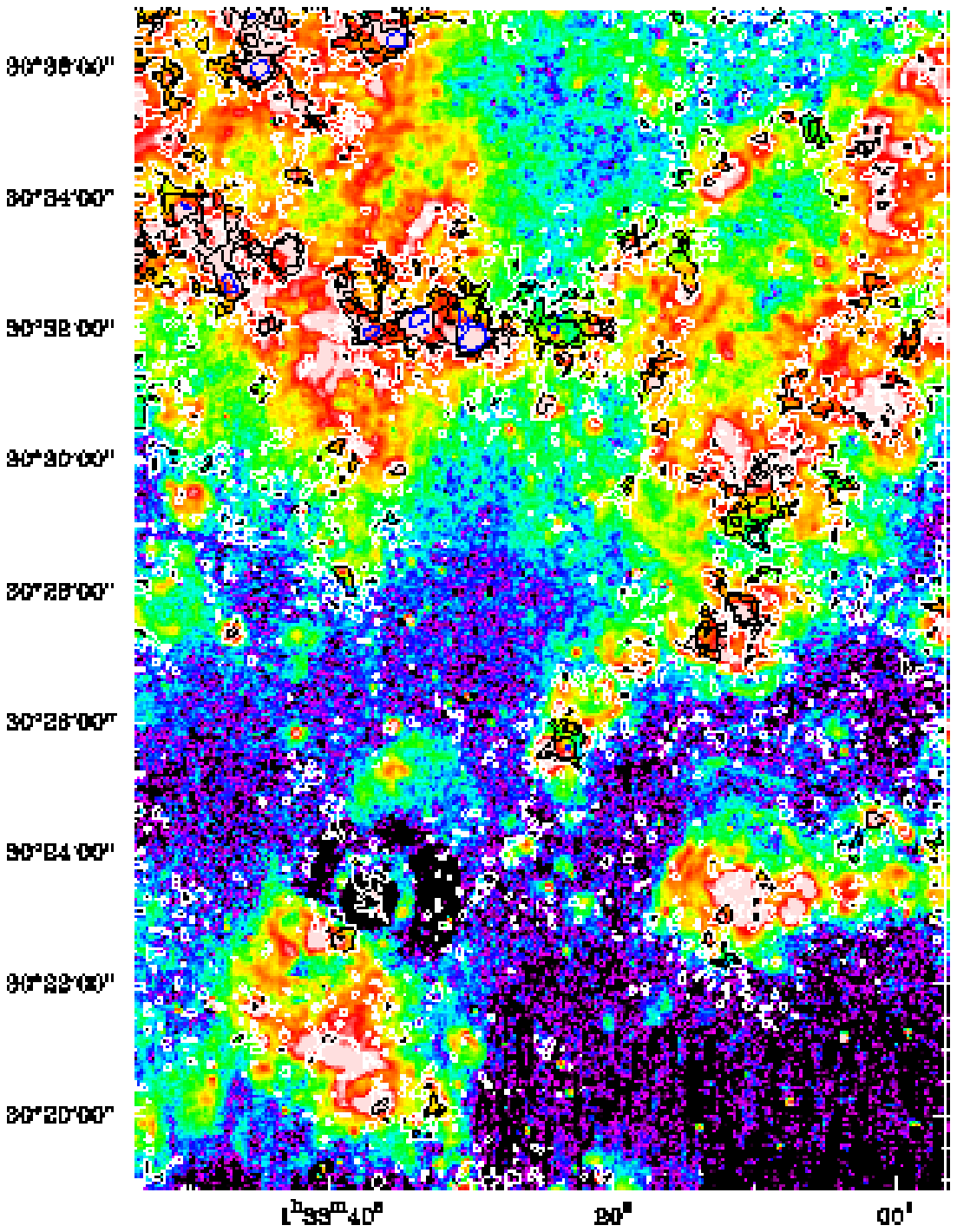}
		\caption{Southern part of KPNO H${\alpha}$ image with \ICO main beam contours of 1~(white), 2,~4~(black), 8~(blue) K$\kms$. The beam size is shown as a white dot in the lower left corner.
} 
	\end{flushleft}
\end{figure*}
}

{ 
\begin{figure*}
	[tbp] 
	\begin{flushleft}
		
		\includegraphics[angle=0,width=18cm]{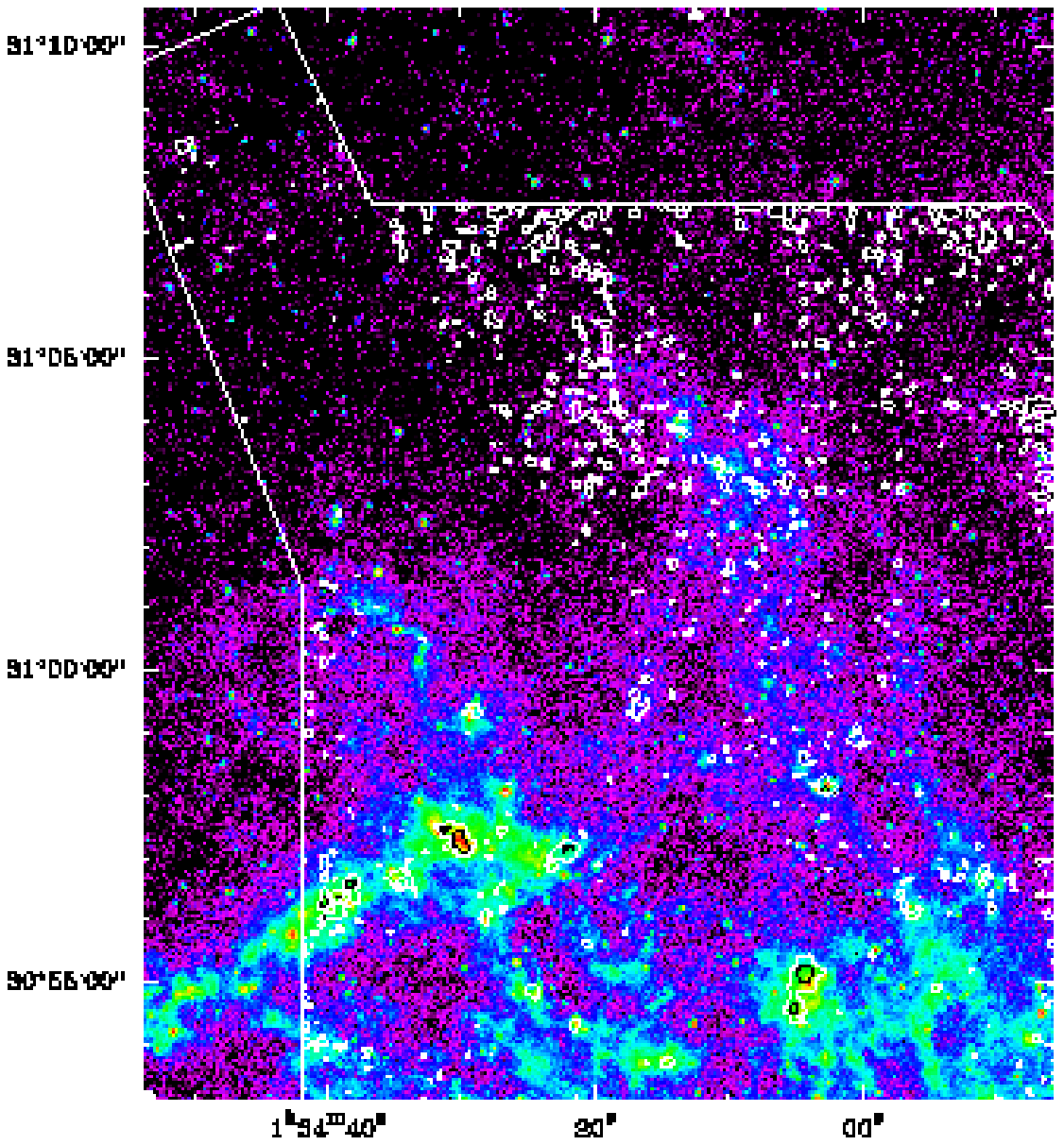}
		\caption{Northern part of \emph{Spitzer} 8$\mum$ image with \ICO main beam contours of 1~(white), 2,~4~(black), 8~(blue) K$\kms$. The beam size is shown as a white dot in the lower left corner.
} 
	\end{flushleft}
\end{figure*}
}

{ 
\begin{figure*}
	[tbp] 
	\begin{flushleft}
		
		\includegraphics[angle=0,width=18cm]{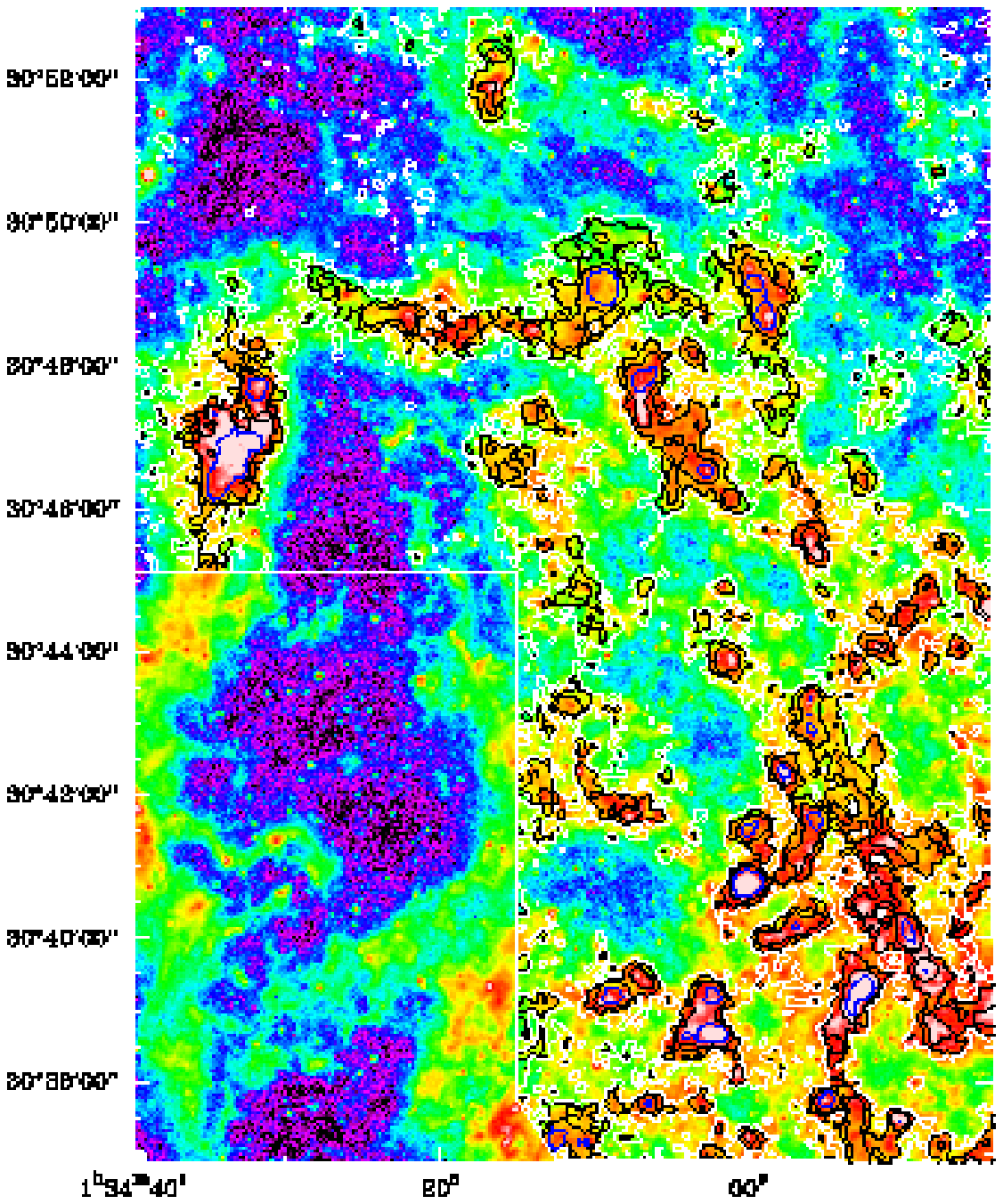}
		\caption{Center part of \emph{Spitzer} 8$\mum$ image with \ICO main beam contours of 1~(white), 2,~4~(black), 8~(blue) K$\kms$. The beam size is shown as a white dot in the lower left corner.
} 
	\end{flushleft}
\end{figure*}
}

{ 
\begin{figure*}
	[tbp] 
	\begin{flushleft}
		
		\includegraphics[angle=0,width=18cm]{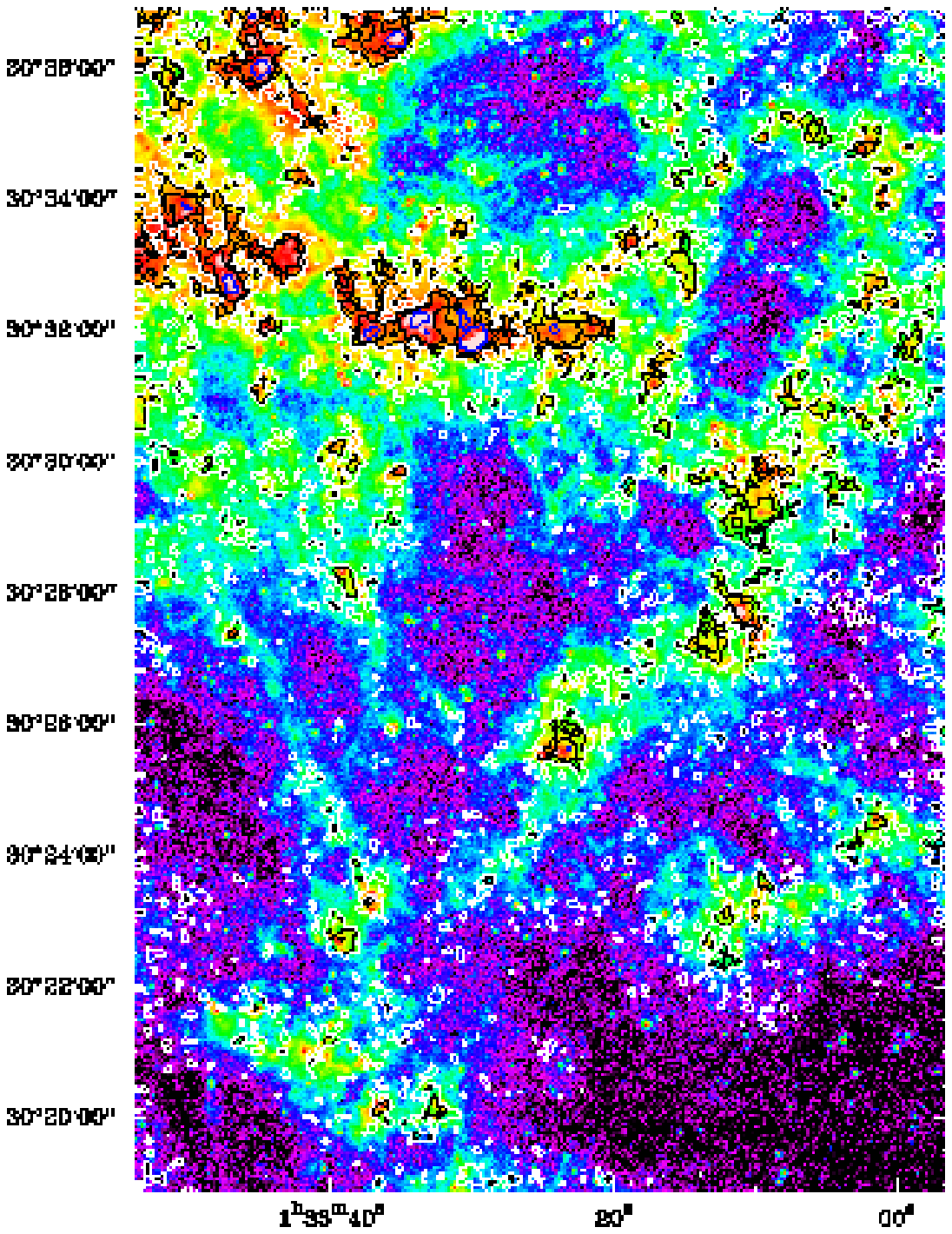}
		\caption{Southern part of \emph{Spitzer} 8$\mum$ image with \ICO main beam contours of 1~(white), 2,~4~(black), 8~(blue) K$\kms$. The beam size is shown as a white dot in the lower left corner.
} 
	\end{flushleft}
\end{figure*}
}

{ 
\begin{figure*}
	[tbp] 
	\begin{flushleft}
		
		\includegraphics[angle=0,width=18cm]{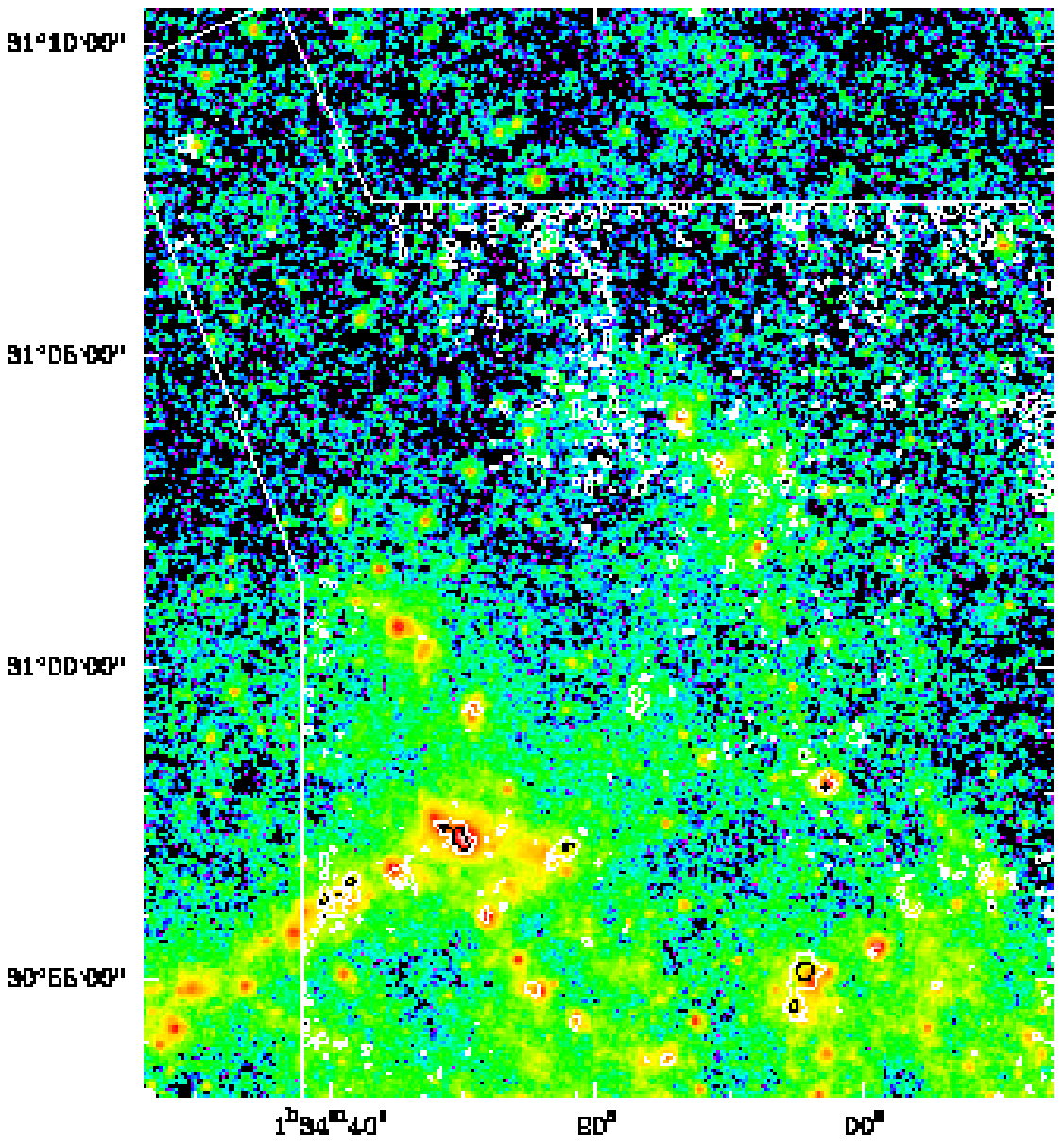}
		\caption{Northern part of \emph{Spitzer} 24$\mum$ image with \ICO main beam contours of 1~(white), 2,~4~(black), 8~(blue) K$\kms$. The beam size is shown as a white dot in the lower left corner.
} 
	\end{flushleft}
\end{figure*}
}

{ 
\begin{figure*}
	[tbp] 
	\begin{flushleft}
		
		\includegraphics[angle=0,width=18cm]{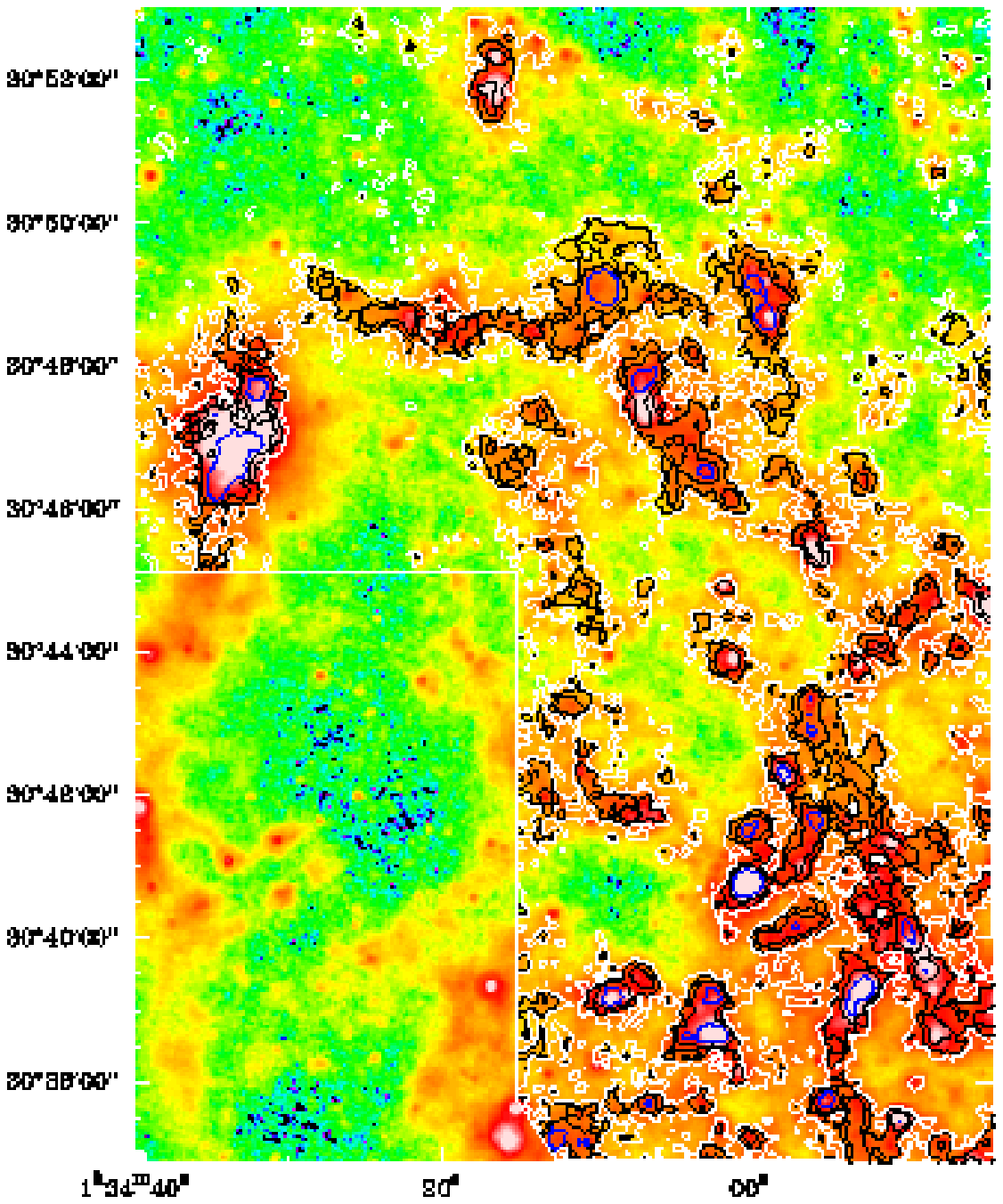}
		\caption{Center part of \emph{Spitzer} 24$\mum$ image with \ICO main beam contours of 1~(white), 2,~4~(black), 8~(blue) K$\kms$. The beam size is shown as a white dot in the lower left corner.
} 
	\end{flushleft}
\end{figure*}
}

\onlfig{9}{ 
\begin{figure*}
	[tbp] 
	\begin{flushleft}
		
		\includegraphics[angle=0,width=18cm]{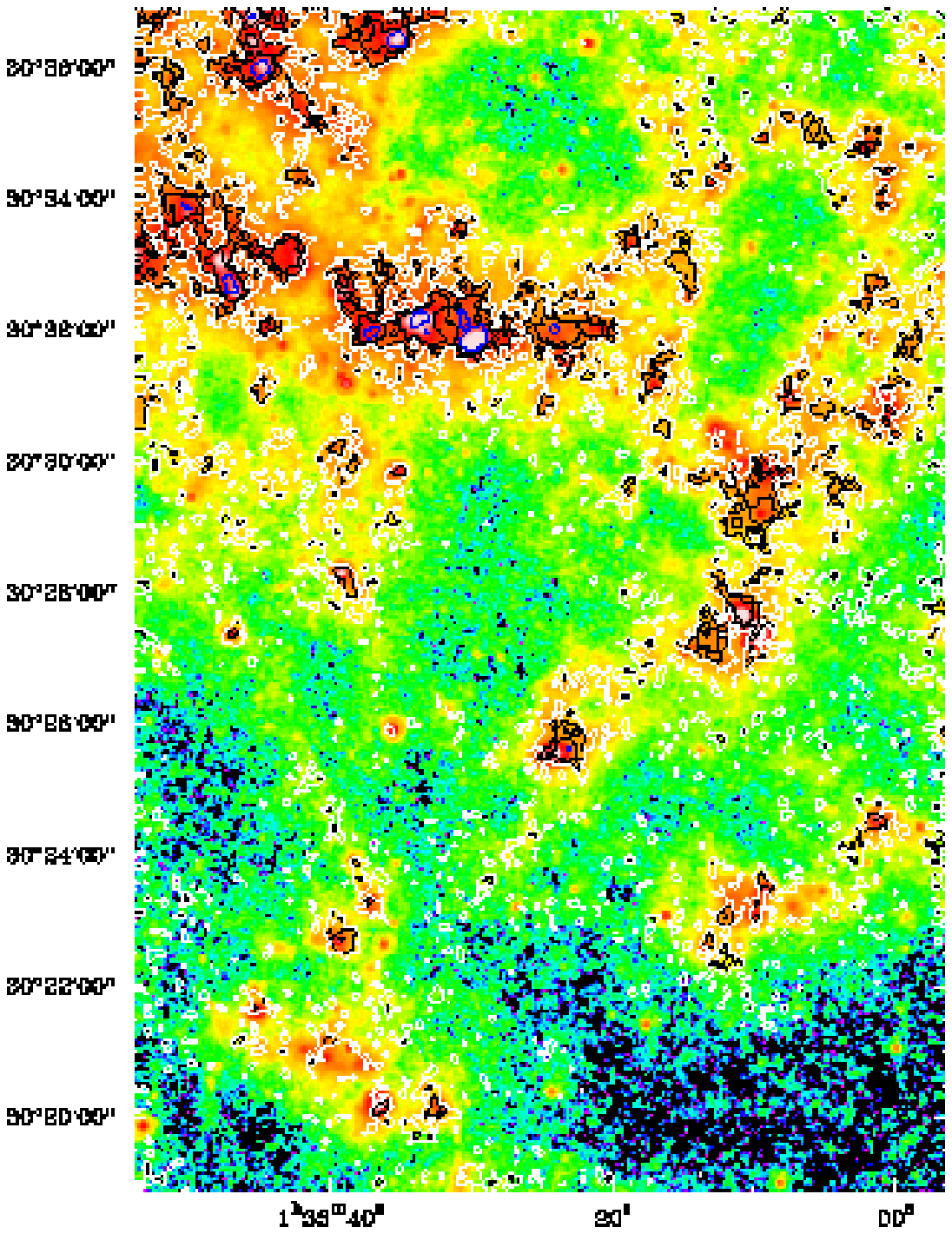}
		\caption{Southern part of \emph{Spitzer} 24$\mum$ image with \ICO main beam contours of 1~(white), 2,~4~(black), 8~(blue) K$\kms$. The beam size is shown as a white dot in the lower left corner.
} 
	\end{flushleft}
\end{figure*}
}

{ 
\begin{figure*}
	[tbp] 
	\begin{flushleft}
		
		\includegraphics[angle=0,width=18cm]{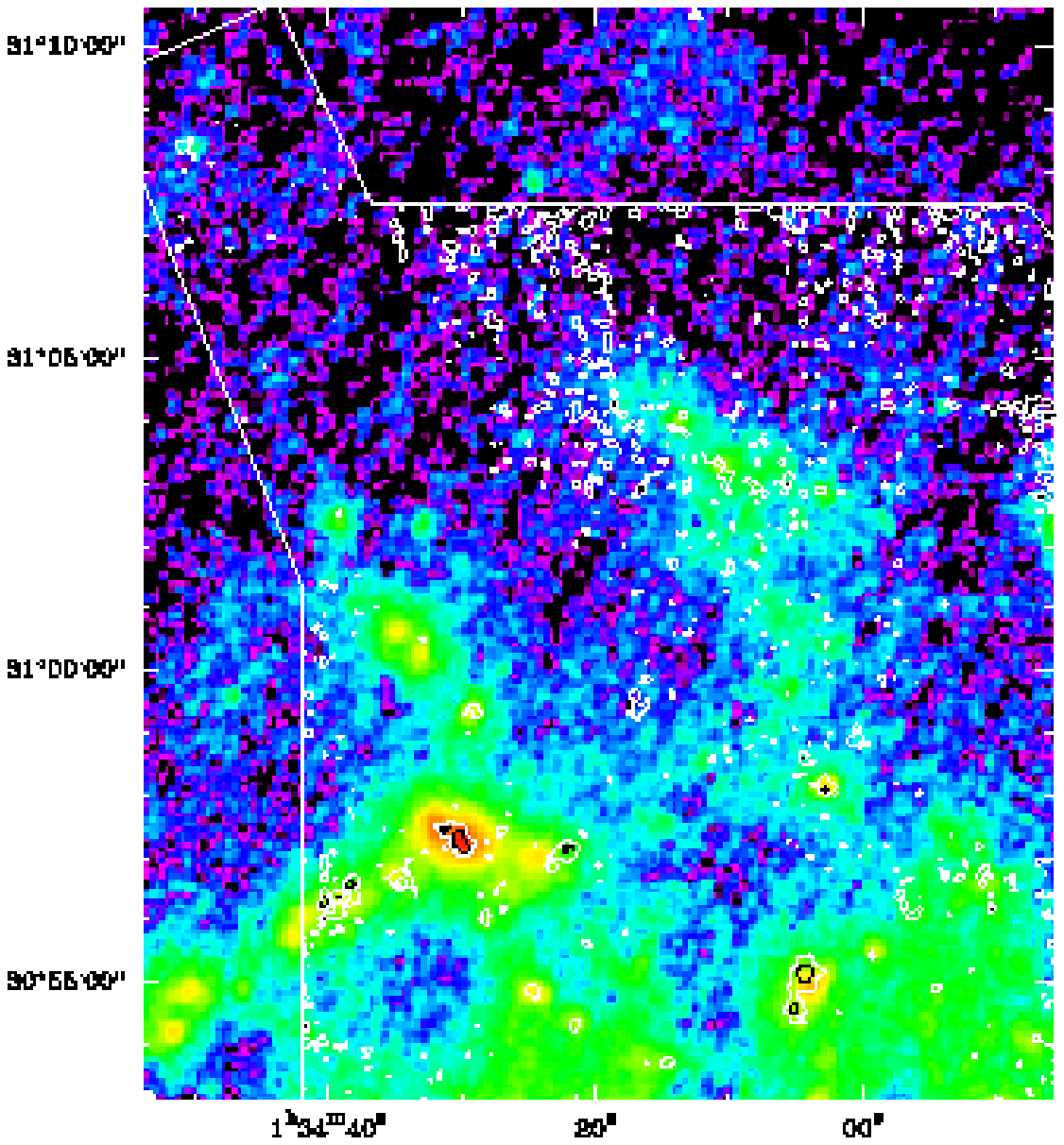}
		\caption{Northern part of \emph{Spitzer} 70$\mum$ image with \ICO main beam contours of 1~(white), 2,~4~(black), 8~(blue) K$\kms$. The beam size is shown as a white dot in the lower left corner.
} 
	\end{flushleft}
\end{figure*}
}

{ 
\begin{figure*}
	[tbp] 
	\begin{flushleft}
		
		\includegraphics[angle=0,width=18cm]{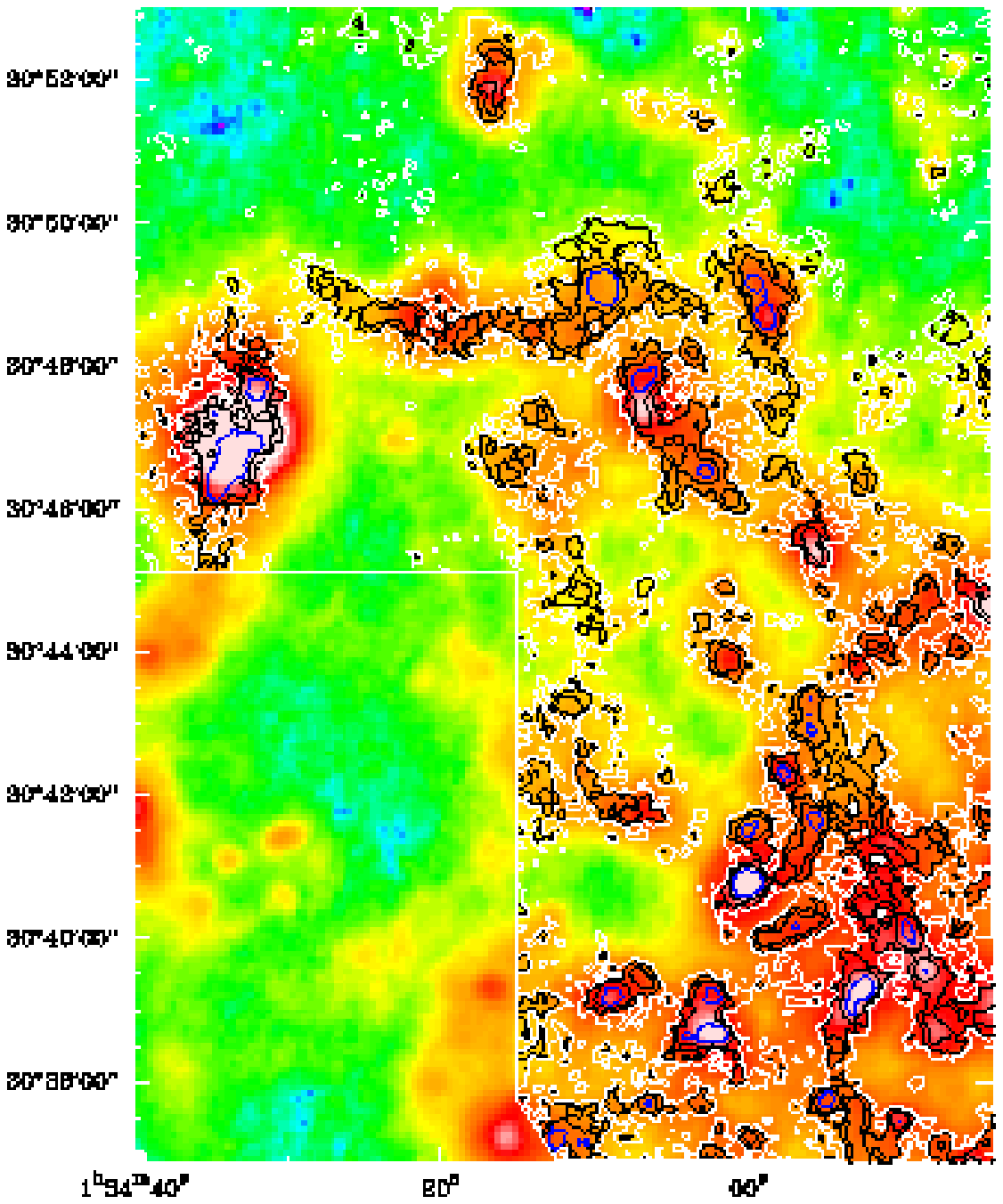}
		\caption{Center part of \emph{Spitzer} 70$\mum$ image with \ICO main beam contours of 1~(white), 2,~4~(black), 8~(blue) K$\kms$. The beam size is shown as a white dot in the lower left corner.
} 
	\end{flushleft}
\end{figure*}
}

{ 
\begin{figure*}
	[tbp] 
	\begin{flushleft}
		
		\includegraphics[angle=0,width=18cm]{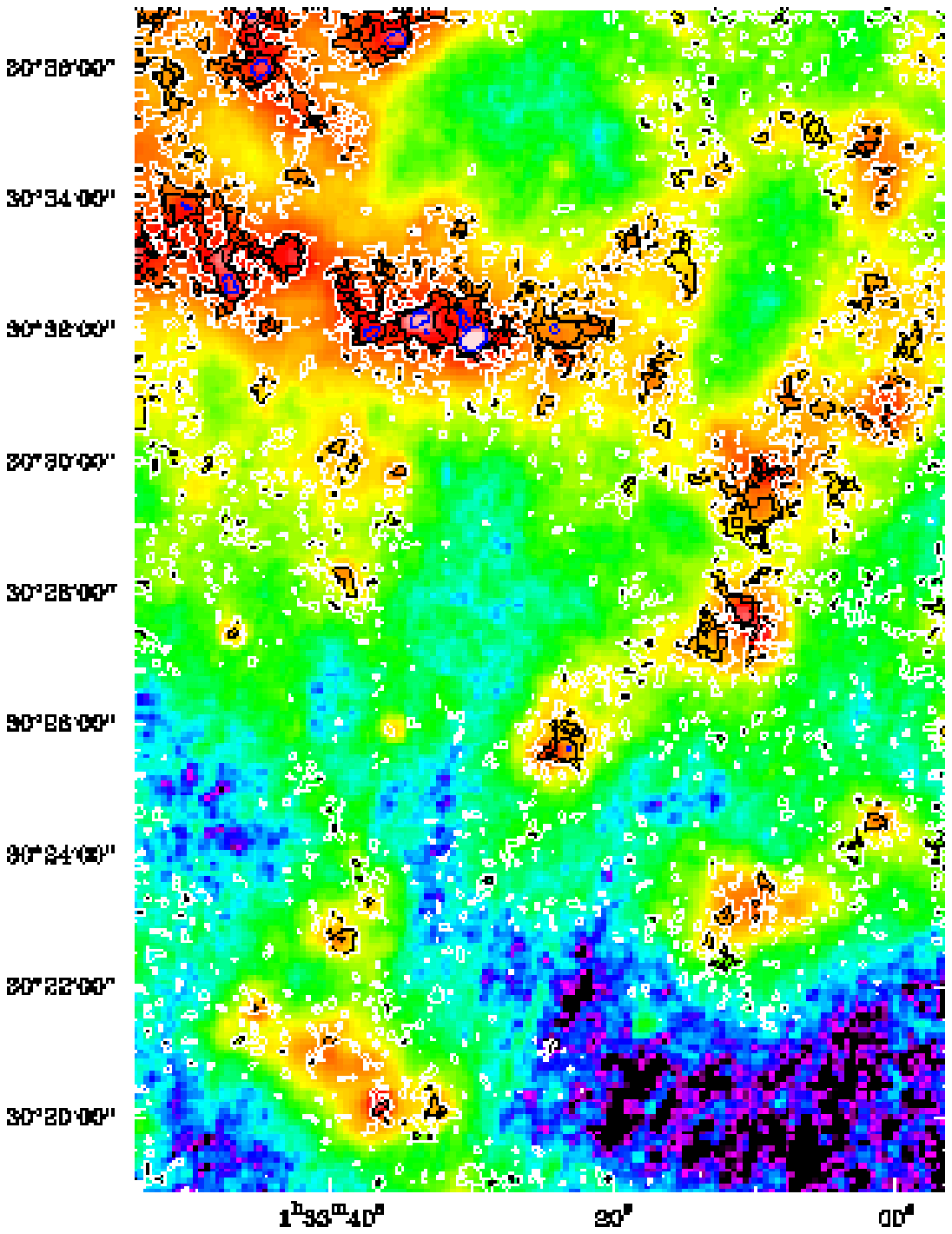}
		\caption{Southern part of \emph{Spitzer} 70$\mum$ image with \ICO main beam contours of 1~(white), 2,~4~(black), 8~(blue) K$\kms$. The beam size is shown as a white dot in the lower left corner.
} 
	\end{flushleft}
\end{figure*}
} 
{ 
\begin{figure*}
	[tbp] 
	\begin{flushleft}
		
		\includegraphics[angle=0,width=18cm]{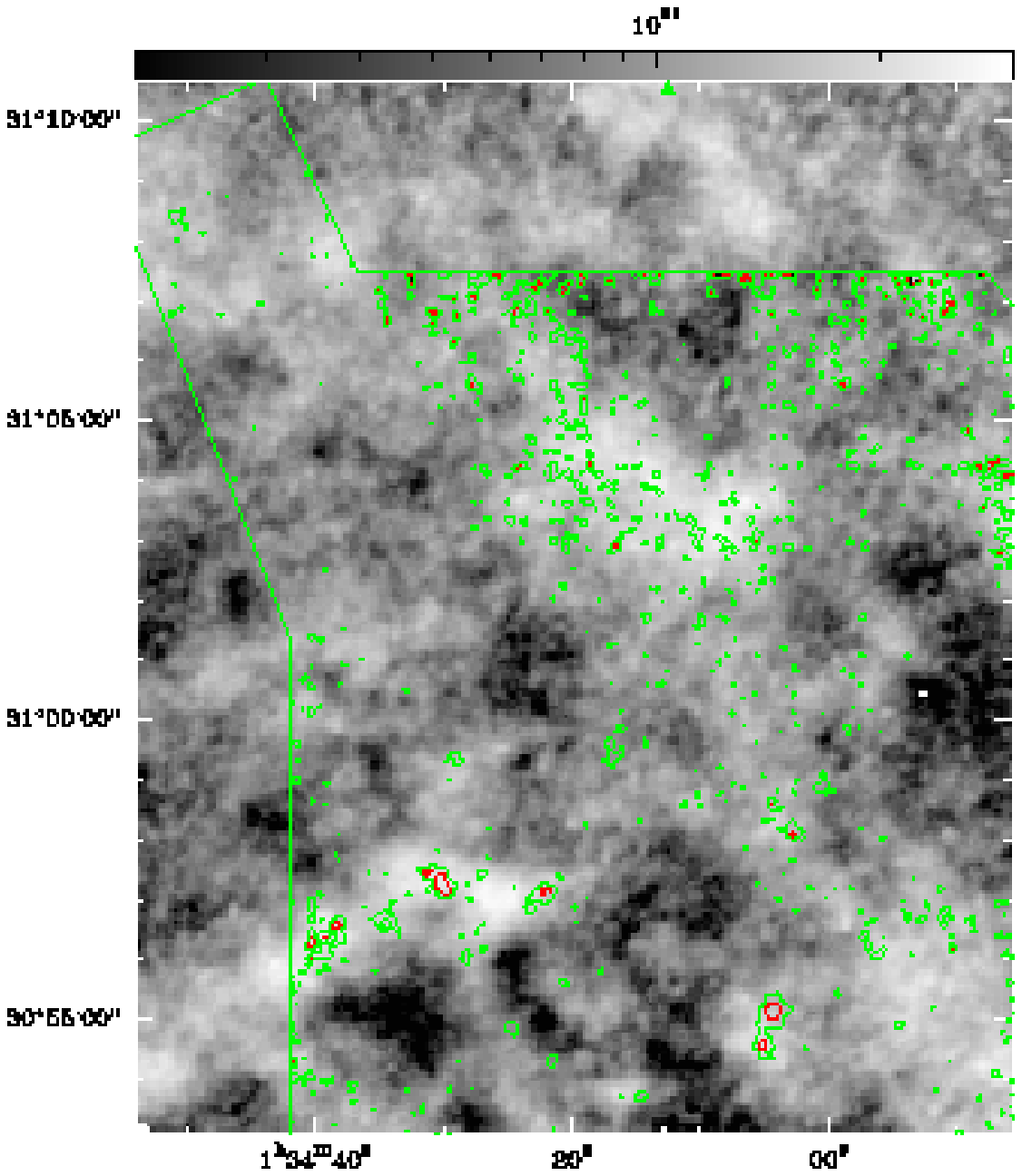}
		\caption{Northern part of VLA \ion{H}{i} 21~cm integrated intensity image with \ICO main beam contours of 1~(green), 2~(red), ~4(black), 8~(blue) K$\kms$. The beam size is shown as a white dot in the lower left corner.
} 
	\end{flushleft}
\end{figure*}
}

{ 
\begin{figure*}
	[tbp] 
	\begin{flushleft}
		
		\includegraphics[angle=0,width=18cm]{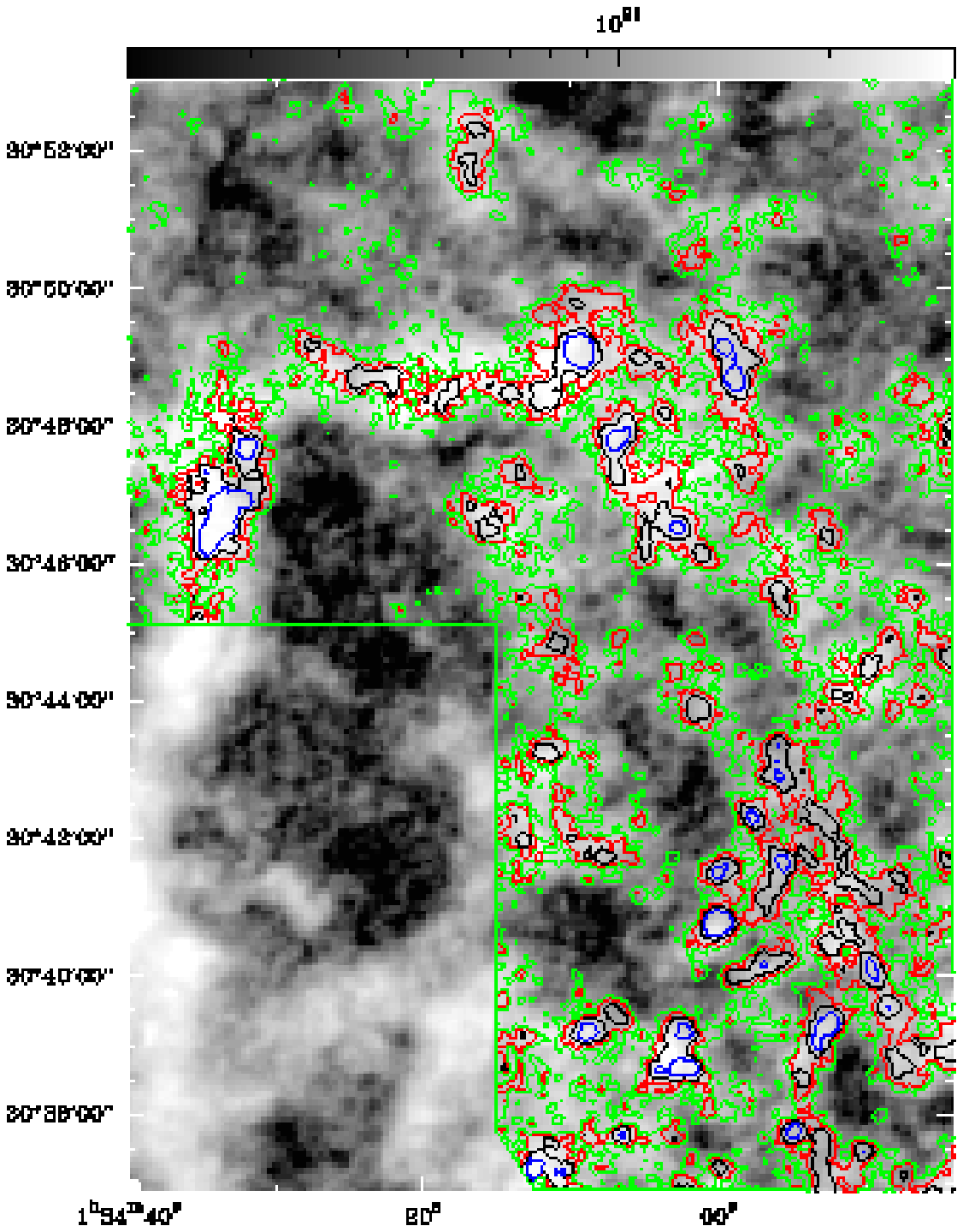}
		\caption{Center part of VLA \ion{H}{i} 21~cm integrated intensity image with \ICO main beam contours of 1~(green), 2~(red), ~4(black), 8~(blue) K$\kms$. The beam size is shown as a white dot in the lower left corner.
} 
	\end{flushleft}
\end{figure*}
}

{ 
\begin{figure*}
	[tbp] 
	\begin{flushleft}
		
		\includegraphics[angle=0,width=18cm]{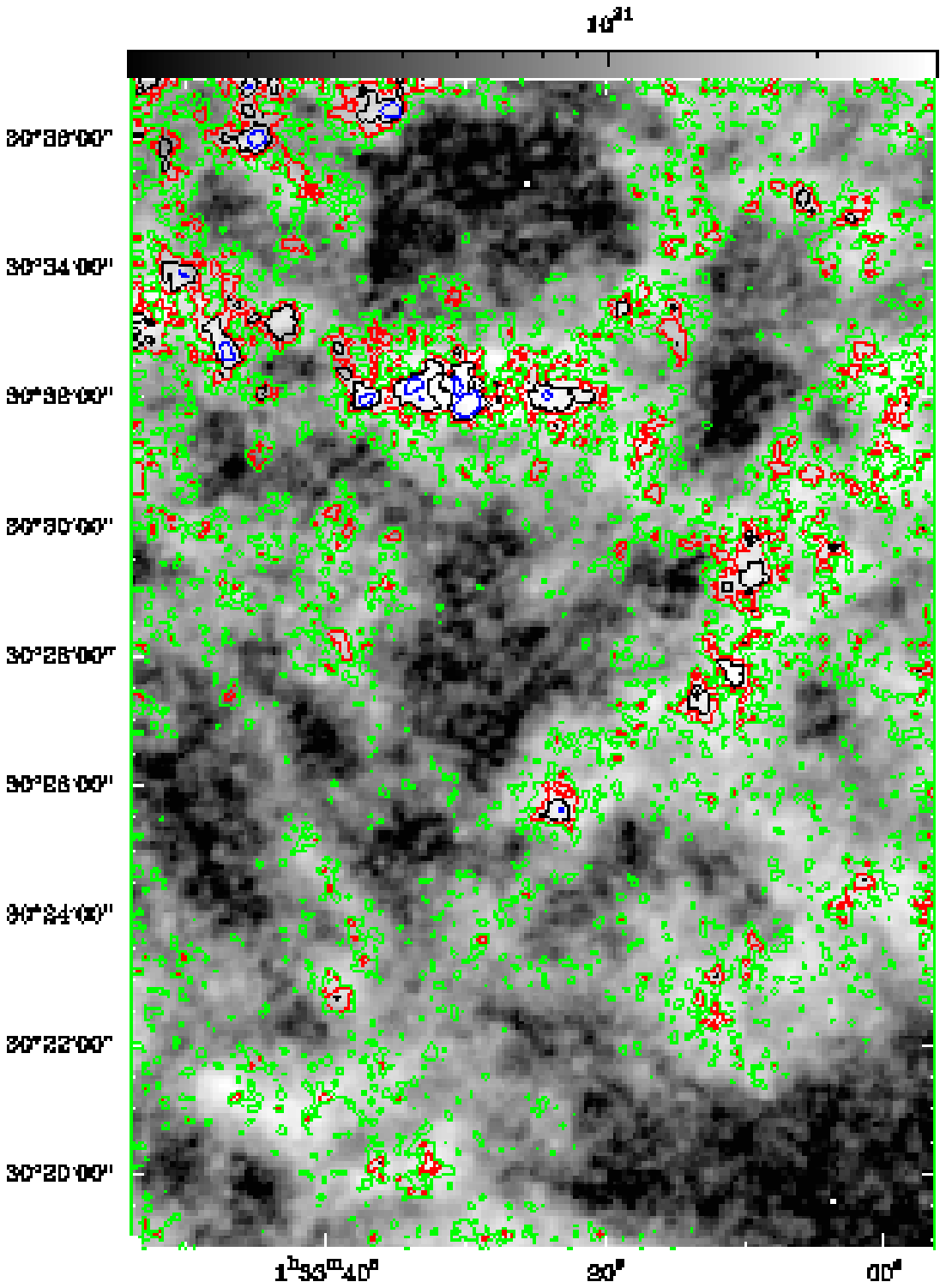}
		\caption{Southern part of VLA \ion{H}{i} 21~cm integrated intensity image with \ICO main beam contours of 1~(green), 2~(red), ~4(black), 8~(blue) K$\kms$. The beam size is shown as a white dot in the lower left corner.
} 
	\end{flushleft}
\end{figure*}
}

\end{appendix}

\end{document}